\documentclass[prd,aps,twocolumn,showpacs,showkeys,nofootinbib,floatfix]{revtex4-2}

\usepackage{amsmath,amssymb}
\usepackage{graphicx}
\usepackage[colorlinks=true,linkcolor=blue,citecolor=blue,urlcolor=blue]{hyperref}
\graphicspath{ {./Figures/} } 

\usepackage{booktabs}
\usepackage{placeins}
\usepackage[normalem]{ulem}
\usepackage{xcolor}

\begin{document}

\title{Probing In-Medium Neutron-to-Dark-Fermion Conversion with Multimessenger Neutron Star Observations}

\author{Davood Rafiei Karkevandi$^{1}$}
\email{davood.rafiei@uwr.edu.pl}

\author{Mahboubeh Shahrbaf$^{1,2}$}
\email{mahboubeh.shahrbafmotlagh@uwr.edu.pl}

\author{Hamid Reza Moshfegh$^{3,4}$}
\email{hmoshfegh@ut.ac.ir}

\author{Reza Karimi$^{3}$}
\email{rezakarimi0225@ut.ac.ir}

\author{Gordon Baym$^{5}$}
\email{gbaym@illinois.edu}

\affiliation{$^1$Institute of Theoretical Physics, University of Wroclaw, 50-204 Wroclaw, Poland}

\affiliation{$^2$Nishina Center for Accelerator-Based Science, RIKEN, Wako 351-0198, Japan}

\affiliation{$^3$Department of Physics, University of Tehran, North Karegar Avenue, Tehran 14395-547, Iran}

\affiliation{$^4$Centro Brasileiro de Pesquisas Fısicas, Rua Dr. Xavier Sigaud,
150, URCA, Rio de Janeiro CEP 22290-180, RJ, Brazil}

\affiliation{$^5$Department of Physics, University of Illinois, 1110 W. Green Street, Urbana, IL 61801, USA}

\date{\today}

\begin{abstract}
Neutron stars  provide a natural environment for probing in-medium neutron-to-dark-fermion conversion and its impact on the dense-matter equation of state (EOS). In this work, we use multimessenger neutron star observations to map the excluded and not-excluded regions of the dark-sector parameter space for  an ensemble of EOSs within a non-linear $\sigma$--$\omega$--$\rho$ relativistic mean-field (RMF) framework
 and determine their overlap across the ensemble. We include a fermionic dark component with repulsive self-interactions, parameterized by a coupling strength $G_\chi$. We consider masses $m_{\chi}$ near neutron mass as well as heavier states accessible in neutron star cores at sufficiently large neutron chemical potentials. To reduce the sensitivity of our results to the choice of nucleonic EOS within the adopted RMF framework, we scan $m_\chi=939$--$1500~\mathrm{MeV}$ and $G_\chi=1$--$100~{\rm fm}^2$ across a 21-EOS nucleonic ensemble, organized into a main baseline set of 13 EOSs together with representative soft and stiff limiting cases. 
 
 The resulting stellar sequences are tested against the requirement that the EOS can support at least $2M_\odot$ mass neutron stars, the  GW170817 tidal-deformability bounds, and NICER mass--radius constraints, where we consider the Illinois--Maryland and X-PSI/Amsterdam inference sets separately.  For the full ensemble of 21 EOSs, we find
that both NICER analyses exclude similar regions of
dark-sector parameter space across the entire ensemble.
These common excluded regions have approximate
minimum--maximum extents of $m_\chi = 939$--$1300$ MeV
and $G_\chi = 1$--$11$ fm$^2$.  Unlike the full 21-EOS ensemble, the 13 main-baseline set has a common not-excluded region whose extent depends 
on the NICER mass--radius analysis considered. The soft and stiff limiting cases show how these regions shift when the nucleonic EOS predicts a maximum neutron star
mass close to the $2M_\odot$ threshold  or predicts large tidal deformabilities. Our results show that, across the 21 calibrated nucleonic RMF EOSs considered here, weak dark-sector repulsion is robustly disfavored across a broad dark-fermion mass range, while the extent of the remaining not-excluded region depends on both
the choice of NICER analysis and the stiffness of the
baryonic EOS.
\end{abstract}


\maketitle

\section{Introduction}
\label{sec:introduction}

Dark matter  constitutes a major component of the Universe, yet
its microscopic nature and interactions with ordinary matter remain unknown \cite{SciPostPhysRev.1}. In recent years, neutron stars  have emerged as promising laboratories for probing dark matter physics across different stages of their evolution \cite{Grippa:2024ach,Brax:2026cmh,Bramante:2023djs}. 
Depending on the model, dark matter may be captured or produced inside neutron stars, potentially leaving observable signatures in their structure and evolution \cite{Karkevandi:2021ygv,Shakeri:2022dwg,RafieiKarkevandi:2026tsh,Konstantinou:2024ynd,Issifu:2025jac,Kunert:2026hmx,Pitz:2024xvh,Cipriani:2025wem,Das:2025fyf,Hajkarim:2024ecp,Rutherford:2024uix,Guha:2024pnn,Avila:2023rzj}. 
A wide range of observations, from bulk properties (masses, radii, and tidal deformabilities) to cooling and heating trends, oscillations and instabilities, and gravitational-wave signals from mergers, has therefore been used to place increasingly strong constraints on dark matter models \cite{Karkevandi:2024vov,Shahrbaf:2025hsw,Liu:2025qco,RafieiKarkevandi:2021hcc,Rather:2026aed,Rutherford:2026dxk,Liu:2024rix,Sedaghat:2026okv,Sedaghat:2025dzt,Shawqi:2024jmk,Issifu:2026diw,Srikanth:2025lic,Routaray:2024fcq,Fakhry:2026cmi,Zhang:2026rkg,Arvikar:2025dwl,Lazarte:2025etl,Harris:2025ewk,Kumaran:2026zbt,Sen:2024yim,Cronin:2023xzc,Diedrichs:2023trk,Lourenco:2025ugk}. 

Among the possible mechanisms for producing dark matter in compact objects \cite{Berezhiani:2020zck,Ellis:2018bkr,Nelson:2018xtr,Allahverdi:2024ygd,McKeen:2021jbh,Khalaf:2026bkv,Heeck:2026hgy,Shahrbaf:2022upc,Shahrbaf:2024gdm,Shahrbaf:2023uxy,Pospelov:2026dnb}, neutrons may couple to a hidden sector through interactions that allow their conversion into final states containing one or more baryon-number-carrying dark fermions \cite{Berryman:2022zic,Berryman:2023rmh}. Such conversion provides a connection between dark matter physics and dense baryonic matter \cite{McKeen:2018xwc,Baym:2018ljz,Husain:2022bxl,Motta:2018rxp,Kopp:2018jom}.  The interiors of neutron stars can reach densities at which the neutron chemical potential, $\mu_n$, exceeds the neutron rest mass by several hundred MeV, and even reaches GeV values in the cores of massive stars, comparable to the possible dark-fermion masses, $m_\chi$, considered in this work.
In such an environment, conversion of neutrons to dark matter particles is governed by chemical equilibrium rather than by vacuum-decay kinematics. 
If a neutron can convert into a dark fermion $\chi$, even weak microscopic couplings can, in principle, drive the system toward equilibrium on astrophysical timescales \cite{universe9110484}.

Crucially, the conversion can proceed in the stellar core once the neutron chemical potential becomes large enough that, to within interaction effects,  $\mu_n \gtrsim m_\chi$. 
As a result, channels that are kinematically forbidden for an isolated neutron, e.g., $m_\chi>m_n$, may become accessible at high densities \cite{McKeen:2018xwc,Baym:2018ljz}. 
This appearance is analogous to that of heavier hadronic degrees of freedom, when the relevant chemical-potential thresholds are reached \cite{Looee:2025dgx,Jangal:2025maa,Karimi:2022vxw, Shahrbaf:2019wex}.
A dark-fermion population, once present, changes the pressure--energy-density relation and therefore impacts macroscopic observables such as the maximum mass, radius, and tidal deformability.  Therefore, neutron stars are potentially powerful probes of GeV-scale dark sectors \cite{RafieiKarkevandi:2026tsh,Biesdorf:2024dor,Issifu:2026ehm,Ivanytskyi:2019wxd}, particularly through neutron-to-dark-fermion conversion scenarios that are inaccessible in terrestrial experiments \cite{McKeen:2018xwc,Baym:2018ljz}.

For dark-fermion masses close to the neutron mass, neutron-to-dark-sector conversion could be relevant to terrestrial experiments, where it has been proposed as a possible explanation of the neutron-lifetime anomaly \cite{Fornal:2018eol,Wietfeldt:2024oku}. Measurements of the neutron lifetime show a persistent tension between the ``bottle'' method, which counts surviving stored ultracold neutrons \cite{Musedinovic:2024gms}, and the ``beam'' method, which infers the lifetime by detecting $\beta$-decay products, typically protons \cite{Yue:2013qrc}. The published UCN$\tau$ average gives
$\tau_n^{\rm bottle}
=877.83\pm0.22~(\mathrm{stat})
{}^{+0.20}_{-0.17}~(\mathrm{syst})~\mathrm{s}$ \cite{Musedinovic:2024gms}, whereas the beam measurement gives
$\tau_n^{\rm beam}
=887.7\pm1.2~(\mathrm{stat})
\pm1.9~(\mathrm{syst})~\mathrm{s}$ \cite{Yue:2013qrc}. Their central values differ by approximately $9.9~{\rm s}$. Although this discrepancy may arise from unresolved experimental systematics, it remains under investigation through independent approaches, including the electron-counting method \cite{Fuwa:2024cdf,Koch:2024cfy}. 

   Notably, the possibility of a discrepancy has motivated several
beyond-Standard-Model explanations involving dark-sector
particles \cite{Cline:2018ami,Berezhiani:2018eds,Fornal:2020gto}.
In particular, the original proposal suggested 
a branching fraction of approximately $1\%$ for exotic
neutron decays involving a dark fermion with a mass
near $938$ MeV
 to explain the neutron lifetime anomaly \cite{Fornal:2018eol}. Several neutron dark-decay modes have also been proposed
\cite{Fornal:2023wji}, among which non-radiative channels
such as $n\rightarrow\chi\phi$ and multiparticle final states
such as $n\rightarrow\chi\chi\chi$ have received particular
attention \cite{Cline:2018ami,Strumia:2021ybk}.  However, experimental searches have placed stringent
constraints on such explanations \cite{Tang:2018eln,Klopf:2019afh,Broussard:2021eyr,UCNA:2018hup,Dubbers:2018kgh}, including a
model-dependent upper bound of order $10^{-5}$ on the
free-neutron dark-decay branching fraction from
${}^{6}\mathrm{He}$ measurements \cite{LeJoubioux:2023usk}.  Nevertheless, dark-decay scenarios proposed in connection with the neutron-lifetime anomaly subsequently motivated a number of studies of neutron-to-dark-sector conversion in neutron stars \cite{universe9110484,Gardner:2023wyl}. Such conversion
in dense matter can occur independently of whether dark
decay explains the anomaly. In particular, experimental
constraints on exotic free-neutron decay do not
automatically rule out conversion in neutron star matter.

Early studies of neutron stars \cite{McKeen:2018xwc,Baym:2018ljz,Motta:2018rxp} showed that if a neutron can convert into  a single weakly interacting dark fermion $\chi$ with baryon number $B_\chi=1$ and mass close to $m_n$, the resulting dark component softens the equation of state (EOS) and typically lowers the maximum supported mass below the observed $\sim 2\,M_\odot$ pulsars \cite{Demorest:2010bx}. 
Therefore, for the single-$\chi$ scenario in neutron star
matter, compatibility with the requirement that the EOS can support two-solar-mass neutron stars generally requires repulsive dark self-interactions, which can be modeled as an effective vector repulsion,  parametrized by an effective coupling $G_\chi$ \cite{Grinstein:2018ptl}. 
Such self-interactions are also independently motivated on galactic scales, where self-interacting dark matter has been widely discussed as a possible way to ease small-scale structure tensions (e.g.\ core--cusp behavior in dwarf galaxies) \cite{Adhikari:2022sbh,Tulin:2017ara}.

Beyond the properties of the dark sector, the impact of
neutron-to-dark-fermion conversion on neutron star observables
also depends on the underlying dense-matter EOS.
Since the true EOS of neutron-star matter remains uncertain,
exploring a range of EOS stiffnesses and nuclear-matter
properties allows us to assess this dependence.
In particular, using nucleonic EOSs helps isolate the effects
of dark-fermion conversion from those of other exotic
degrees of freedom. Moreover, adopting EOSs derived from
the same theoretical framework allows for the examination of
these effects under controlled variations of nuclear-matter
properties. Nevertheless, the existence of neutron stars with masses
of about $2M_\odot$ provides a robust observational
requirement with which any physically viable EOS must be consistent, regardless of its microscopic
physics. This provides a general constraint on
neutron-to-dark-fermion conversion, particularly at low
dark-fermion masses and weak self-interaction strengths,
where the resulting EOS softening can significantly
reduce the maximum supported mass.

  In the neutron star context, most previous work has focused on
the single-$\chi$ scenario with a dark-fermion mass
close to the neutron mass ($m_\chi \approx m_n$),
originally motivated by the neutron-lifetime anomaly,
and typically included repulsive dark self-interactions
to remain compatible with massive neutron stars
\cite{Harris:2025isp,Shirke:2023ktu,Shirke:2024ymc,
Thakur:2024btu,Das:2025pjl,Vikiaris:2026ofd,Routaray:2024lni,Husain:2023fwb,Shirke:2025lsu,Arvikar:2026hhi,Keung:2020teb,Divaris:2026vow,Thakur:2025zhi}.
The multi-particle scenario $n\to 3\chi$ (with fractional baryon number per particle) has received comparatively less attention and is on occasion discussed without requiring strong self-interactions \cite{Strumia:2021ybk,Husain:2022brl,Darini:2022mcy}. Finally, the possibility of an in-medium channel to a heavier dark fermion ($m_\chi>m_n$) was highlighted in the pioneering neutron star studies of \citet{McKeen:2018xwc} and \citet{Baym:2018ljz}. These studies, as well as the few more recent works on this regime \cite{Harris:2025isp,Kalita:2026sgu,Heeck:2026hgy}, have explored different dark-fermion masses and hadronic
EOS models, but have not systematically investigated
the joint mass-interaction strength $(m_\chi,G_\chi)$ parameter space for dark
fermions with repulsive self-interactions across a range of EOSs derived within the same theoretical framework. 
These limitations motivate our systematic scan over $(m_\chi,G_\chi)$ combined with a nucleonic EOS ensemble representing soft, intermediate, and stiff behavior. Using different multimessenger constraints, this strategy allows us to identify excluded and not-excluded regions that are common across the set of nucleonic EOSs we consider.

In this work, we study the single-$\chi$ neutron-conversion channel in
neutron star matter, imposing the chemical-equilibrium condition
$\mu_\chi=\mu_n$ and allowing for repulsive dark self-interactions.
We consider both the region near the neutron-mass scale and heavier
dark fermions with $m_\chi>m_n$ that can become accessible only in dense
matter.

To constrain this scenario, we confront the resulting stellar sequences with three complementary observational inputs: the requirement that the EOS support the observed $\sim 2\,M_\odot$ neutron stars, the tidal-deformability limit from GW170817, and the NICER telescope mass--radius inferences. For NICER, we consider the independent Illinois--Maryland and X-PSI/Amsterdam mass--radius inferences separately within the same analysis framework.

To model neutron star matter, we employ relativistic mean-field (RMF) theory within a non-linear $\sigma$--$\omega$--$\rho$ framework, which provides a thermodynamically consistent description of dense nucleonic matter and can be naturally extended to include additional fermionic degrees of freedom. To reduce the sensitivity of our results to the choice of nucleonic EOS within the RMF framework, we adopt an ensemble of \emph{21 unified} (crust-to-core) RMF EOSs \cite{Malik:2024nva}, selected from a larger RMF family whose couplings were constrained in a Bayesian analysis incorporating nuclear-physics and astrophysical constraints \cite{malik:2023mnx}. This ensemble approach allows us to assess how the inferred dark-sector constraints vary among the nucleonic EOSs considered. Given the uncertainties in the EOS of matter at
supranuclear densities, the differences among the
21 nucleonic EOSs considered here reflect variations
in the parameters of the adopted RMF framework and
should not be interpreted as a quantitative estimate
of the uncertainty in the true EOS of neutron-star matter.

Considering purely nucleonic
 EOSs allows us
to isolate the effects of
neutron--dark-fermion conversion from those associated with changes in the high-density
composition. Under this
assumption, the inferred dark-sector constraints do not
include modifications of the EOS and stellar structure
caused by non-nucleonic components.
Nevertheless, degrees of freedom beyond nucleons may
emerge in neutron star cores at sufficiently high densities \cite{Baym:2019yyo,Baym:2017whm}.   This possibility is particularly relevant for massive neutron stars:
the NICER radius inference for the $2.08\,M_\odot$ pulsar
PSR J0740+6620 has been examined within the QHC21
framework, which incorporates a crossover from hadronic
to quark matter at high densities \cite{Kojo:2021wax}. Therefore, extending the present analysis to EOSs incorporating such additional high-density physics is an important direction for future work.

This paper is organized as follows. In Sec.~\ref{sec:model}, we describe the RMF framework used for the nucleonic sector and its extension to include neutron-to-dark-fermion conversion with repulsive dark self-interactions. In Sec.~\ref{sec:analysis_framework}, we introduce the nucleonic EOS ensemble and the dark-sector parameter ranges, and present the stellar
structure results for purely nucleonic stars together with the adopted multimessenger constraints. In Sec.~\ref{sec:eos_composition_effects}, we examine the effects of the dark-sector parameters on the EOS, composition, and stellar observables. Section~\ref{sec:constraint_strategy} defines the constraint strategy and notation used in the subsequent analyses. Sections~\ref{sec:main_baseline_results}--\ref{sec:stiff_results} present the separate Illinois--Maryland and X-PSI/Amsterdam analyses for the main-baseline, soft, and stiff EOS groups. In Sec.~\ref{sec:full_ensemble_results}, we compare the EOS-overlapped conservative exclusions obtained from the two NICER inference sets for the complete 21-EOS ensemble. Finally, Sec.~\ref{sec:conclusions} summarizes our main conclusions regarding
dark-sector constraints and their dependence on the nucleonic EOS. 

\section{Theoretical framework}
\label{sec:model}

We work in natural units $\hbar=c=k_B=1$ and consider cold, catalyzed neutron star matter at $T=0$ under charge neutrality and $\beta-$equilibrium. 
The stellar EOS is constructed by combining a nucleonic RMF baseline supplemented by a dark matter component generated via
neutron-to-dark-fermion conversion and subject to repulsive dark
self-interactions. 
The nucleonic sector is described within relativistic mean-field (RMF) theory using a non-linear $\sigma$--$\omega$--$\rho$ model: nucleons interact through the isoscalar-scalar field $\sigma$, the isoscalar-vector field $\omega_\mu$, and the isovector-vector field $\vec{\rho}_\mu$. Additional non-linear mesonic
self-interaction terms are included in order to reproduce
empirical properties of nuclear matter and to control the
stiffness of the EOS at supranuclear densities.
We work in the mean-field approximation for uniform matter, where the meson fields are replaced by their expectation values, so nucleons propagate as quasiparticles with an effective mass (set by $\sigma$) and chemical potentials shifted by the vector mean fields ($\omega_0$ and $\rho_{0,3}$). 
The standard model matter composition consists of neutrons, protons,
electrons, and muons in weak equilibrium.

\subsection{Nucleonic sector: non-linear RMF}
\label{sec:rmf_nucleonic}

The hadronic baseline is described within a non-linear relativistic mean-field (RMF) model with $\sigma$, $\omega_\mu$, and $\vec{\rho}_\mu$ mesons.
The hadronic sector Lagrangian density is
\begin{equation}
\mathcal{L}=\mathcal{L}_N+\mathcal{L}_M+\mathcal{L}_{\rm NL}+\mathcal{L}_{\ell},
\end{equation}
where the nucleonic part reads
\begin{equation}
\mathcal{L}_N
=\bar{\Psi}\Big[\gamma^\mu\big(i\partial_\mu-g_\omega \omega_\mu-\tfrac{1}{2}g_\rho\,\vec{\tau}\!\cdot\!\vec{\rho}_\mu\big)
-(m_N-g_\sigma\sigma)\Big]\Psi,
\end{equation}
with $\Psi=(p,n)^T$ the nucleon isodoublet, $m_N$ the bare nucleon mass, and $\vec{\tau}$ the Pauli matrices in isospin space.
The meson contribution is
\begin{align}
\mathcal{L}_M
&=\frac{1}{2}\partial_\mu\sigma\,\partial^\mu\sigma-\frac{1}{2}m_\sigma^2\sigma^2
-\frac{1}{4}\omega_{\mu\nu}\omega^{\mu\nu}+\frac{1}{2}m_\omega^2\omega_\mu\omega^\mu \nonumber\\
&\quad
-\frac{1}{4}\vec{\rho}_{\mu\nu}\!\cdot\!\vec{\rho}^{\,\mu\nu}
+\frac{1}{2}m_\rho^2\,\vec{\rho}_\mu\!\cdot\!\vec{\rho}^{\,\mu}.
\end{align}
with field tensors $\omega_{\mu\nu}=\partial_\mu\omega_\nu-\partial_\nu\omega_\mu$ and
$\vec{\rho}_{\mu\nu}=\partial_\mu\vec{\rho}_\nu-\partial_\nu\vec{\rho}_\mu$.
Non-linear meson self-interactions and the mixed $\omega$--$\rho$ term are included through
\begin{equation}
\begin{aligned}
\mathcal{L}_{\rm NL}
&=-\frac{b}{3}\,m_N\,(g_\sigma\sigma)^3
  -\frac{c}{4}\,(g_\sigma\sigma)^4 \\
&\quad
  +\frac{\xi}{4!}\,\bigl(g_\omega^2\,\omega_\mu\omega^\mu\bigr)^2
  +\Lambda_\omega\,\bigl(g_\omega^2\omega_\mu\omega^\mu\bigr)\,
                  \bigl(g_\rho^2\,\vec{\rho}_\nu\!\cdot\!\vec{\rho}^{\,\nu}\bigr).
\end{aligned}
\end{equation}

The constants $g_\sigma$, $g_\omega$, and $g_\rho$ denote the nucleon--meson couplings of the $\sigma$, $\omega_\mu$, and $\vec{\rho}_\mu$ fields, respectively. The non-linear couplings $b$, $c$, $\xi$, and $\Lambda_\omega$ control distinct aspects of the density dependence of the RMF EOS. 
The cubic and quartic $\sigma$ self-interactions ($b$ and $c$) are introduced to tune the incompressibility around saturation of symmetric nuclear matter, while the $\omega^4$ term (parameter $\xi$) regulates the high-density repulsion and therefore the stiffness of the EOS at supranuclear density (larger $\xi$ generally softens the EOS at high density). 
The mixed $\omega$--$\rho$ coupling $\Lambda_\omega$ primarily controls the density dependence of the isovector channel and, in practice, allows one to vary the symmetry-energy slope at saturation. Different calibrated choices of these couplings generate the range of
high-density stiffness and symmetry-energy behavior represented by the
nucleonic EOS ensemble introduced in Sec.~\ref{sec:analysis_framework}.

Leptons are modeled as free relativistic Fermi gases, with
\begin{equation}
\mathcal{L}_{\ell}=\sum_{\ell=e,\mu}\bar{\psi}_\ell\left(i\gamma^\mu\partial_\mu-m_\ell\right)\psi_\ell .
\end{equation}

\subsection{Mean-field equations and chemical potentials}
\label{sec:rmf_mf}

In the mean-field approximation, we replace the meson fields by their
ground-state expectation values. For uniform matter, the scalar field is
spatially constant, while only the time-like components of the vector fields
remain nonzero:
\begin{equation}
\omega^\mu \rightarrow \delta^{\mu 0}\omega_0,\qquad
\vec{\rho}^{\,\mu}\rightarrow
\delta^{\mu 0}\rho_{0,3}\hat{e}_3 .
\end{equation}
The nucleons then move independently in these classical background fields.
The nucleon effective mass is
\begin{equation}
m_N^\ast = m_N-g_\sigma\sigma .
\end{equation}
For each fermion species $i = n,p$ with Fermi momentum $k_{F,i}$, the number density is
\begin{equation}
n_i=\frac{k_{F,i}^3}{3\pi^2},
\end{equation}
and the nucleon scalar density is
\begin{equation}
n_i^{s}=\frac{1}{\pi^2}\int_0^{k_{F,i}}\frac{m_N^\ast}{\sqrt{k^2+m_N^{\ast 2}}}\,k^2\,dk.
\end{equation}

The meson mean fields are determined from the Euler--Lagrange equations:
\begin{align}
m_\sigma^2\sigma + b\,m_N\,g_\sigma^3\sigma^2 + c\,g_\sigma^4\sigma^3 &= g_\sigma\,(n_n^s+n_p^s), \label{eq:sigma_mf_final}\\
m_\omega^2\omega_0 + \frac{\xi}{6}\,g_\omega^4\omega_0^3 + 2\Lambda_\omega g_\omega^2g_\rho^2\,\omega_0\,\rho_{0,3}^2 &= g_\omega\,(n_n+n_p), \label{eq:omega_mf_final}\\
m_\rho^2\rho_{0,3} + 2\Lambda_\omega g_\omega^2g_\rho^2\,\omega_0^2\,\rho_{0,3} &= \frac{1}{2}g_\rho\,(n_p-n_n). \label{eq:rho_mf_final}
\end{align}
The nucleon chemical potentials are
\begin{align}
\mu_n &= \sqrt{k_{F,n}^2+m_N^{\ast 2}} + g_\omega\omega_0 - \frac{1}{2}g_\rho\rho_{0,3}, \\
\mu_p &= \sqrt{k_{F,p}^2+m_N^{\ast 2}} + g_\omega\omega_0 + \frac{1}{2}g_\rho\rho_{0,3},
\end{align}
while for leptons $\mu_\ell=\sqrt{k_{F,\ell}^2+m_\ell^2}$.

\subsection{Neutron-to-dark-fermion conversion and self-interactions}
\label{sec:dm_sector}

Following the thermodynamic treatment of
Refs.~\cite{Baym:2018ljz,McKeen:2018xwc}, we include a fermionic dark state $\chi$ that can be populated via
neutron-to-dark-fermion conversion processes in dense neutron star matter,
focusing on the single-$\chi$ channel ($n\leftrightarrow\chi$ ).
In practice, the conversion is treated through chemical-equilibrium
conditions rather than by tracking microscopic reaction rates, which is
sufficient for determining the equilibrium composition and EOS.

The dark fermion is assumed to carry baryon number $B_\chi=1$, so that the conversion of neutron to dark particle is implemented through the chemical-equilibrium condition
\begin{equation}
\mu_\chi=\mu_n .
\label{eq:chemical_equilibrium}
\end{equation}
This framework covers two physically relevant mass regimes: dark-fermion masses close to the neutron mass, $m_\chi\simeq m_n$, and heavier states with $m_\chi>m_n$. In the latter case, the in-medium conversion channel can become accessible inside neutron star cores, where the neutron chemical potential increases at high density.

The conserved baryon density is then defined as
\begin{equation}
n_B = n_n+n_p+n_\chi ,
\end{equation}
where $n_n$, $n_p$, and $n_\chi$ are the neutron, proton, and dark-fermion number densities, respectively. 
Repulsive dark self-interactions are included through an effective vector interaction in the dark sector. 
This is motivated by the fact that a weakly interacting dark fermion populated through neutron conversion can strongly soften the EOS, while a repulsive dark interaction increases the pressure of the dark component and can help maintain compatibility with massive neutron stars. 
We model this interaction by introducing a vector mediator $V_\mu$ coupled to the dark-fermion current,
\begin{equation}
\begin{aligned}
\mathcal{L}_\chi
&=\bar{\chi}\left(i\gamma^\mu\partial_\mu-m_\chi\right)\chi
- g_V\,\bar{\chi}\gamma^\mu\chi\,V_\mu \\
&\quad
-\frac{1}{4}V_{\mu\nu}V^{\mu\nu}
+\frac{1}{2}m_V^2 V_\mu V^\mu ,
\end{aligned}
\end{equation}
where $g_V$ is the dark vector coupling, $m_V$ is the mediator mass, and
$V_{\mu\nu}=\partial_\mu V_\nu-\partial_\nu V_\mu$ is the corresponding field-strength tensor.

Applying the same mean-field prescription to the dark vector mediator gives
\begin{equation}
V^\mu \rightarrow \delta^{\mu 0}V_0 .
\end{equation} 
The field equation for the mediator then gives
\begin{equation}
m_V^2 V_0 = g_V n_\chi ,
\end{equation}
so that the dark-fermion chemical potential becomes
\begin{equation}
\mu_\chi =
\sqrt{k_{F,\chi}^2+m_\chi^2}
+ g_V V_0
=
\sqrt{k_{F,\chi}^2+m_\chi^2}
+ G_\chi n_\chi .
\label{eq:muchi_final}
\end{equation}
 
The dark-fermion number density and the effective self-interaction strength are defined as
\begin{equation}
n_\chi=\frac{k_{F,\chi}^3}{3\pi^2},
\qquad
G_\chi \equiv \left(\frac{g_V}{m_V}\right)^2 .
\end{equation}
 In the numerical calculations we express \(G_\chi\) in units of \({\rm fm}^2\).
Larger values of $G_\chi$ correspond to stronger repulsion among dark fermions, which increases the dark-sector contribution to the pressure and reduces the EOS softening caused by neutron-to-dark-fermion conversion.

The neutron-level formulation adopted here should be understood as a low-energy hadronic description. At a more fundamental level, the dark sector is expected to couple to quarks, and microscopic neutron-portal models show how the effective neutron--dark-fermion interaction can arise from quark-level operators \cite{Fornal:2018eol,McKeen:2020oyr,Berryman:2022zic,Girmohanta:2026eqg}. Such an interaction can also operate above the QCD transition in the early Universe, before hadrons are formed \cite{McKeen:2020oyr}. More generally, chemical equilibrium can be expressed in terms of the baryon chemical potential $\mu_\chi=\mu_B$ \cite{Baym:2018ljz}, which, for the purely nucleonic matter considered in the present work, reduces to $\mu_\chi=\mu_n$. The present calculation is restricted to the hadronic phase; extending it to a large quark-matter core would require specifying the corresponding quark-level interaction and treating the dark-sector equilibrium in that phase.

We also emphasize that the repulsive vector self-interaction considered here is only one possible choice for the dark sector. Both attractive and repulsive $\chi$--$\chi$ interactions, as well as interactions between the dark and neutron sectors, have been studied in the neutron-dark-decay context \cite{Grinstein:2018ptl}. Another example combines scalar-mediated dark self-interactions with an effective repulsive neutron--dark matter interaction \cite{Bastero-Gil:2024kjo}. Related compact-star studies, although based on different dark-matter production mechanisms, also illustrate the competing effects of scalar attraction and vector repulsion and show that more general attractive interactions can modify the onset of the new component \cite{Grippa:2024sfu,Heeck:2026hgy}. In the present work, we focus on a repulsive vector $\chi$--$\chi$ interaction as a specific, physically motivated case, leaving attractive, mixed scalar--vector, and additional dark-matter--baryon interactions for future work.

\subsection{Composition: $\beta$ equilibrium and charge neutrality}
\label{sec:equilibrium}

The composition at fixed conserved baryon density $n_B$ is obtained by solving 
Eqs.~\eqref{eq:sigma_mf_final}--\eqref{eq:rho_mf_final} together with the standard conditions of $\beta$ equilibrium and charge neutrality,
\begin{equation}
\mu_n=\mu_p+\mu_e,\qquad 
\mu_e=\mu_\mu,\qquad 
n_p=n_e+n_\mu .
\label{eq:betaeq_final}
\end{equation}
In addition, neutron-to-dark-fermion conversion imposes
the chemical-equilibrium condition given in
Eq.~\eqref{eq:chemical_equilibrium}. 
Solving these equations yields the full set of Fermi momenta $\{k_{F,i}\}$ and mean fields 
$\{\sigma,\omega_0,\rho_{0,3}\}$ at each density.

For $m_\chi>m_n$, $\chi$ is absent at low density and appears only once the neutron chemical potential is large enough to satisfy equilibrium with a nonzero dark density. 
Operationally, we first solve the nucleon--lepton system with $n_\chi=0$ at each $n_B$ and evaluate $\mu_n$. 
The onset of the dark component occurs when
\begin{equation}
\mu_n \geq \mu_\chi(n_\chi\to 0)=m_\chi .
\label{eq:dark_onset_final}
\end{equation}
Above this threshold, $n_\chi$ is determined self-consistently from $\mu_\chi=\mu_n$ together with $\beta$ equilibrium and charge neutrality.

\subsection{Energy density and pressure}
\label{sec:eos_final}

The total energy density is written as
\begin{equation}
\varepsilon
=\sum_{i=n,p,e,\mu,\chi}\varepsilon_i^{\rm kin}
+\varepsilon_{\rm meson}
+\varepsilon_{\chi{\rm -int}},
\end{equation}
where each fermionic kinetic term is
\begin{equation}
\varepsilon_i^{\rm kin}
=\frac{1}{\pi^2}\int_0^{k_{F,i}}\sqrt{k^2+m_i^{\ast 2}}\,k^2\,dk,
\end{equation}
with $m_{n,p}^\ast=m_N^\ast$ and $m_{e,\mu,\chi}^\ast=m_{e,\mu,\chi}$.
The meson-field contribution is
\begin{equation}
\begin{aligned}
\varepsilon_{\rm meson}
&=\frac{1}{2}m_\sigma^2\sigma^2
+\frac{b}{3}m_N(g_\sigma\sigma)^3
+\frac{c}{4}(g_\sigma\sigma)^4
+\frac{1}{2}m_\omega^2\omega_0^2 \\
&\quad
+\frac{\xi}{8}(g_\omega\omega_0)^4
+\frac{1}{2}m_\rho^2\rho_{0,3}^2
+3\Lambda_\omega\,(g_\omega g_\rho \omega_0\rho_{0,3})^2  ,
\end{aligned}
\end{equation}
and the dark repulsion adds
\begin{equation}
\varepsilon_{\chi{\rm -int}}=\frac{1}{2}G_\chi n_\chi^2.
\end{equation}
We compute the pressure at zero temperature using the thermodynamically consistent identity
\begin{equation}
P=\sum_{i=n,p,e,\mu,\chi}\mu_i n_i-\varepsilon.
\label{eq:P_thermo_final}
\end{equation}


\section{Analysis methodology }
\label{sec:analysis_framework}

\subsection{Nucleonic EOS ensemble }
\label{sec:eos_ensemble_final}

To reduce the dependence of our results on any single parametrization within the RMF framework considered here, we employ the published set of \emph{21 unified crust-to-core} nucleonic RMF EOSs constructed within the same non-linear $\sigma$--$\omega$--$\rho$ framework. 
This ensemble was selected to cover a broad range of values of the symmetry-energy slope at saturation, approximately
$22 \lesssim L_{\rm sym,0} \lesssim 70~\mathrm{MeV}$, and therefore probes a broad range of isovector behaviors of dense nucleonic matter. Variations of the nonlinear vector coupling $\xi$ also produce distinct high-density stiffnesses across the ensemble.
 This is particularly important for the present study, since the density dependence of the symmetry energy influences the proton fraction, crust--core transition, radii, and tidal deformabilities, while the high-density stiffness directly affects maximum-mass support. This ensemble therefore allows us to examine how variations within the nucleonic RMF family, particularly in the isovector sector associated with the density dependence of the symmetry energy, affect the inferred dark-sector constraints. It also captures differences in high-density stiffness, allowing us to assess their effect on the maximum mass supported by each EOS.

The EOSs provide a unified crust-to-core description of cold, charge-neutral, $\beta$-equilibrated stellar matter. 
The outer crust is described by the BSk22 equation of
state, based on a Brussels--Montreal Skyrme-type
nuclear energy-density functional \cite{Pearson2018BSk}, while the inner crust is constructed within a compressible liquid-drop approximation and matched consistently to the uniform nucleonic core composed of neutrons, protons, electrons, and muons \cite{Malik:2024nva}. 
The RMF couplings were selected from a broader non-linear RMF parameter family constrained through a Bayesian analysis, incorporating empirical nuclear saturation properties, low-density pure-neutron-matter information from chiral effective-field theory, and basic astrophysical requirements, including the existence of neutron stars with masses above $2\,M_\odot$ \cite{malik:2023mnx}.

However, the variation across these EOSs reflects differences in the parameter values within this RMF model and should not be taken as a quantitative estimate of the uncertainty in the physical EOS. At the same time, the ensemble covers a broad range of soft, intermediate, and stiff EOS behavior, allowing us to test whether the inferred constraints remain consistent across these different nucleonic EOSs. Therefore, a result that is common across all 21 EOSs, or across part of the ensemble, indicates that it is insensitive to the parameter variations represented within this RMF family, rather than independent of the underlying many-body framework. 

This nucleonic composition is a deliberate modeling choice that allows the effects of neutron-to-dark-fermion conversion to be isolated; it should not be interpreted as a claim that the cores of massive neutron stars are necessarily purely nucleonic. Additional high-density degrees of freedom, such as hyperons \cite{Looee:2025dgx,Karimi:2022vxw,Jangal:2025maa}, deconfined quark matter \cite{Kojo:2021wax,Baym:2019iky,Shahrbaf:2021cjz,Ayriyan:2025rub,Shahrbaf:2020uau}, or other exotic states \cite{Shahrbaf:2022upc,Shahrbaf:2023uxy}, may emerge. In particular, QHC21 relates the NICER information on PSR~J0740+6620 to a density regime in which hadronic matter may be evolving toward quark degrees of freedom. Moreover, \citet{Baym:2018ljz} formulated dark-fermion equilibrium through the baryon chemical potential and applied it to QHC18 \cite{Baym:2017whm}, showing that the thermodynamic treatment of conversion can be extended beyond purely nucleonic matter. A qualitatively similar feedback was found in our recent sexaquark study, where the appearance of the sexaquark component in the hadronic phase modified the hadronic EOS used as input to a subsequent hadron--quark crossover construction \cite{Shahrbaf:2024gdm}. A quantitative extension of the present self-interacting model to QHC21, however, would require the crossover EOS to be reconstructed self-consistently if dark conversion modifies the hadronic input, rather than simply replacing the nucleonic EOS with the unmodified QHC21 relation. Such an extension is left for future work.  

In the present work, we divide the published ensemble into three groups of
nucleonic baselines: a 13-EOS main baseline set, representative stiff EOSs,
and the softest EOSs in the ensemble. The main-baseline EOSs are selected because, in the purely nucleonic limit, they satisfy the adopted maximum-mass and tidal-deformability constraints and are broadly compatible with the NICER mass--radius inferences. The stiff cases generally predict larger
radii and tidal deformabilities, whereas the soft cases support maximum masses
close to $2M_\odot$. 
By comparing these EOS groups, we can quantify the dependence
of the inferred dark-sector exclusions on nucleonic stiffness
and identify the common excluded region in the
$(m_\chi,G_\chi)$ plane across the EOS variations represented
within the considered RMF family. The remaining not-excluded parameter space should therefore be
interpreted only within the present multimessenger constraints
and the adopted EOS ensemble, rather than as a uniquely favored or universally
allowed region of the dark-sector parameter space.

\subsection{Scan ranges}
\label{sec:scan_ranges}

 We perform a dense grid scan in the dark-sector parameter
space.  For the dark-fermion mass, we include a single
near-threshold point at $m_\chi=939~\mathrm{MeV}$, close to
the neutron mass, together with a uniform
$5~\mathrm{MeV}$ spacing from $945$ to $1500~\mathrm{MeV}$.  This gives
$N_{m_\chi}=113$ mass points. For the repulsive self-interaction strength, we sample
\begin{equation}
G_\chi = 1,2,3,\ldots,50,55,60,\ldots,95,100 ~{\rm fm}^2,
\end{equation}
corresponding to $N_{G_\chi}=60$ coupling values. Therefore, for each nucleonic
baseline we construct
\begin{equation}
N_{m_\chi}N_{G_\chi}=6780
\end{equation}
dark-matter-admixed EOSs and the corresponding stellar sequences.

Applying this scan to the 13 main baseline EOSs
yields a total of $88140$ models. The same grid is also used for the
representative stiff and soft EOS groups in order to assess how the inferred
dark-sector exclusions depend on the underlying nucleonic stiffness.

\subsection{Reference nucleonic models  and observational constraints}
\label{subsec:nucleonic_reference}

In this subsection, we examine the behavior of our nucleonic EOSs in the absence of a dark component and compare them with the observational constraints that will be used in the dark-sector scan. This purely nucleonic limit, \(n_\chi=0\), provides the reference point for understanding how neutron-to-dark-fermion conversion modifies the stellar sequences. It also clarifies which EOSs are already broadly compatible with the adopted constraints, which ones are relatively stiff and mainly challenged by mass--radius or tidal-deformability bounds, and which ones are sufficiently soft that additional dark-sector softening may endanger the maximum-mass constraint.

The NICER X-ray telescope provides one of the main observational inputs to
our analysis through mass--radius constraints
\cite{Watts:2016uzu,10.1117/12.2231304}, inferred independently by the
X-PSI/Amsterdam and Illinois--Maryland groups
\cite{Rutherford:2024srk,Bogdanov:2021yip}.

The two groups provide independent mass–radius inferences for the same four pulsars: PSR J0030+0451 \cite{Vinciguerra:2023qxq,Miller:2019cac}, PSR J0740+6620 \cite{Salmi:2024aum,Dittmann:2024mbo}, PSR J0437–4715 \cite{Choudhury:2024xbk,Miller:2025qfq} and the most recent and compact source PSR J0614–3329 \cite{Mauviard:2025dmd,Miller:2026vpr}. For each source the first and second references correspond to the X-PSI/Amsterdam and Illinois–Maryland analyses, respectively. We consider the
reported 68\% and 95\% credible mass--radius regions. The X-PSI/Amsterdam and Illinois--Maryland inference sets are applied separately within the same analysis framework.

We also use the tidal-deformability constraint from GW170817, expressed through the adopted bound on \(\Lambda_{1.4}\) \cite{LIGOScientific:2018cki}. The existence of neutron stars with masses around \(2M_\odot\) is used as an additional constraint in the dark-sector scan \cite{Antoniadis:2013pzd}. In the purely nucleonic limit considered in this section, the selected EOSs already satisfy the two-solar-mass requirement. This condition nevertheless remains an important exclusion criterion in the dark-sector scan, because neutron-to-dark-fermion conversion can soften the EOS and reduce the maximum supported mass.

The reference comparison also motivates the EOS grouping used in the
following analysis. We take the subset of 13 EOSs as our main nucleonic
baseline because, already in the purely nucleonic limit and independently
of the dark-sector extension, they provide the most balanced description
among the 21 EOSs considered here. These EOSs support massive neutron stars \cite{Romani:2022jhd}, satisfy the adopted tidal-deformability bound, and predict
mass--radius sequences that are broadly compatible with the NICER
credible regions. In particular, for the Illinois--Maryland inference set, all 13 EOSs intersect the 68\% credible regions of all four pulsars. For the X-PSI/Amsterdam inference set, all 13 EOSs intersect the 68\% credible regions of PSR J0030+0451 and PSR J0740+6620 and the 95\% credible region of PSR J0437--4715, while most also intersect the 95\% credible region of the compact source PSR J0614--3329. For these reasons, this set provides the primary nucleonic reference for constraining the dark-sector parameter space.

The main baseline EOSs avoid the two limiting behaviors represented by
the very stiff and very soft EOSs. The stiff subset tends to predict
larger radii and larger tidal deformabilities, and is therefore mainly
challenged by constraints favoring relatively compact neutron stars
and by the tidal-deformability bound. In this case,
neutron-to-dark-fermion conversion can reduce the radius and tidal
deformability and may move the stellar sequences closer to
observationally favored regions. The soft subset, on the other hand,
predicts maximum
masses close to the 
$2M_\odot$ threshold and is less favored by
 observations suggesting the possible existence of more massive neutron stars \cite{Romani:2025ytn}. Moreover, very soft nucleonic baselines leave limited room for
additional degrees of freedom, whether standard-model hadronic degrees
of freedom such as hyperons or more exotic components such as dark
fermions, because these additional components generally soften the EOS
further and can more easily reduce the maximum supported mass below the
massive-star requirement. Comparing the main baseline, stiff, and soft
subsets allows us to determine excluded and not-excluded regions in the
$(m_\chi,G_\chi)$ plane for each EOS class, while using the main
baseline set as the primary reference for the dark-sector constraints.
This comparison provides the context for interpreting the constraint analyses presented in Secs.~\ref{sec:main_baseline_results}--\ref{sec:full_ensemble_results} and their dependence on the underlying nucleonic EOS.

\begin{figure}[!t]
    \centering
    \includegraphics[width=3.3in]{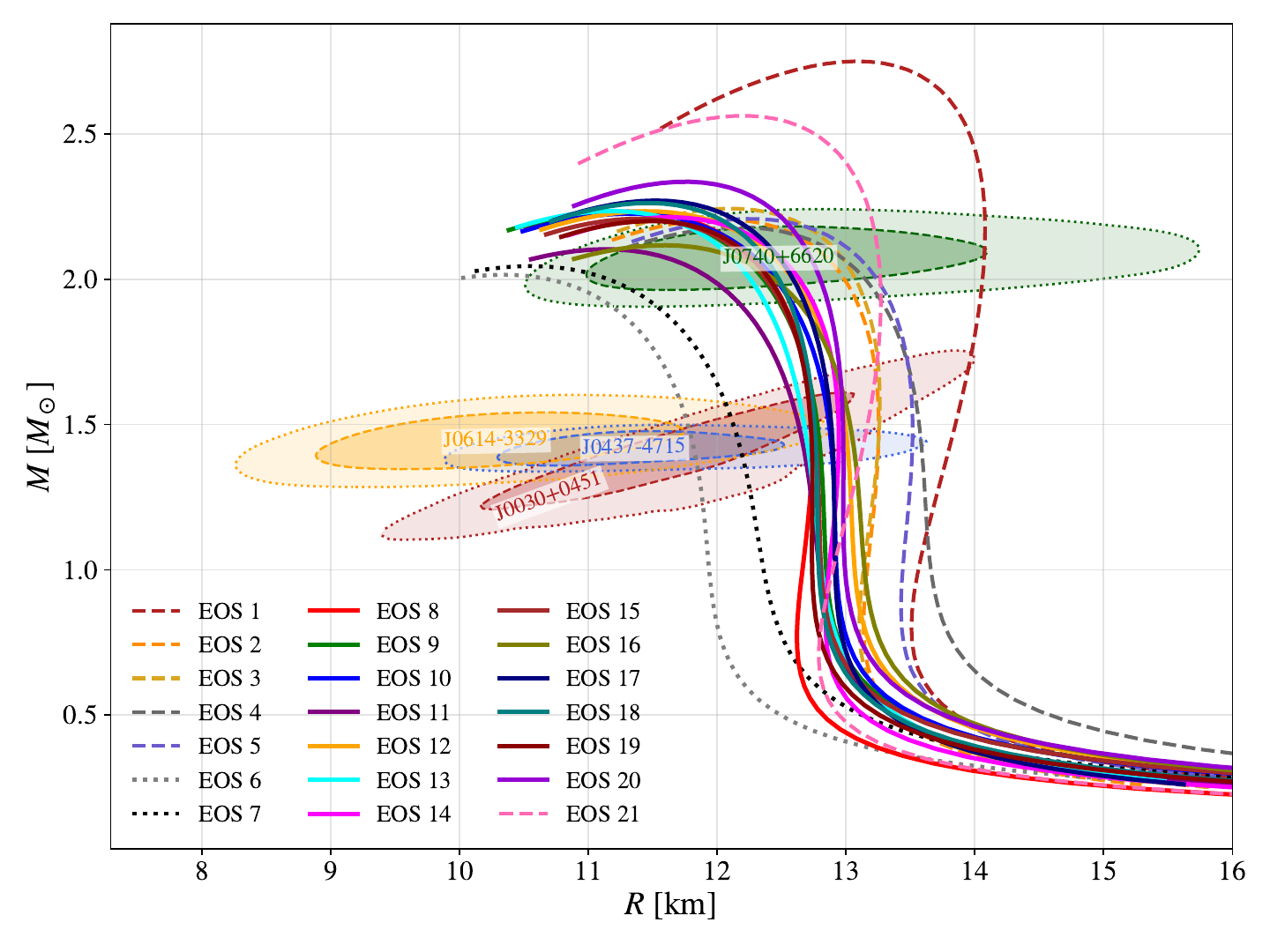}
    \includegraphics[width=3.3in]{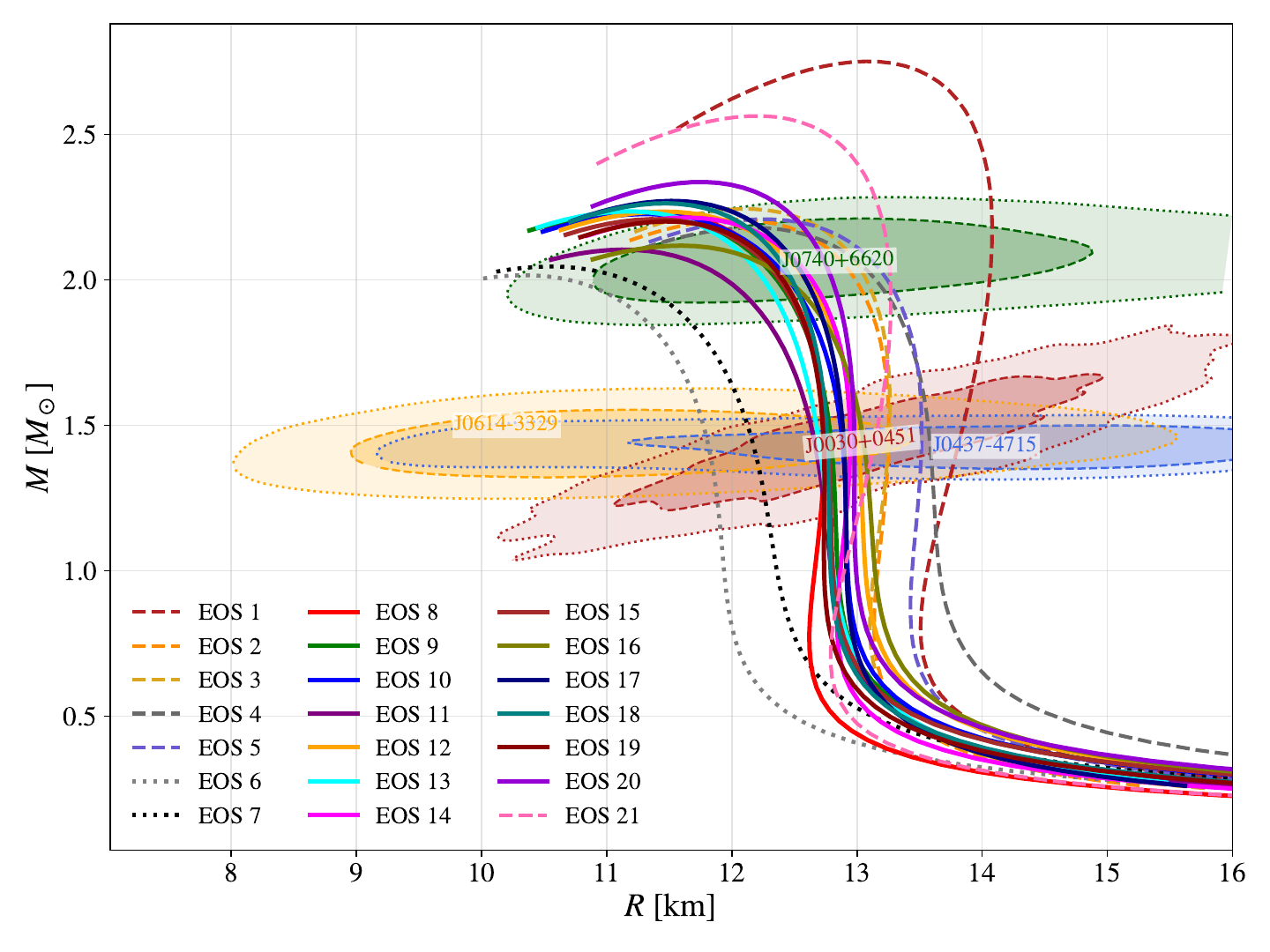}
     \includegraphics[width=3.3in]{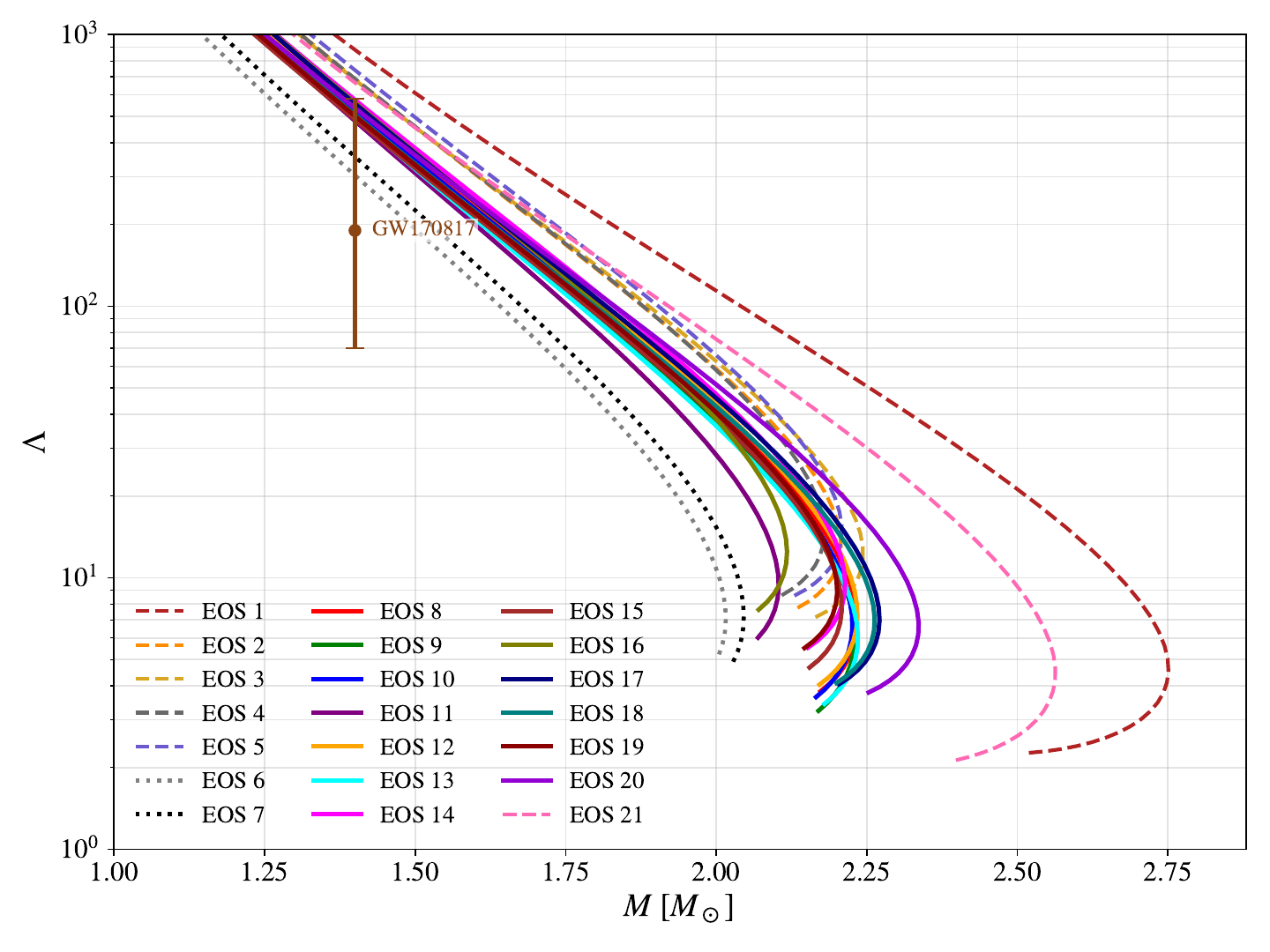}
    \caption{
Purely nucleonic reference sequences for the 21 EOSs considered
in this work, shown for $n_\chi=0$. The top and middle panels compare
the mass--radius sequences with the NICER credible regions from the X-PSI/Amsterdam and Illinois--Maryland analyses, respectively. The bottom panel
shows the corresponding dimensionless tidal deformability as a function
of mass and the adopted GW170817 bound on $\Lambda_{1.4}$. The line styles
indicate the EOS classification used in the dark-sector analysis.
}
\label{fig:nucleonic_reference}
\end{figure}

Figure~\ref{fig:nucleonic_reference} shows the reference behavior of the
selected EOSs in the purely nucleonic limit, $n_\chi=0$. The top and
middle panels compare the mass--radius sequences with the NICER credible
regions obtained by the  X-PSI/Amsterdam and Illinois--Maryland groups,
respectively. The bottom panel compares the corresponding
tidal-deformability sequences with the adopted GW170817 bound on
$\Lambda_{1.4}$. The NICER constraints test the combined mass--radius
behavior of each EOS, rather than only the radius at a fixed mass. A stellar sequence is considered compatible with a given NICER inference if its stable branch intersects the corresponding two-dimensional mass–radius credible region. 

The line styles indicate the EOS classification used in the dark-sector analysis: solid colored curves correspond to the main baseline EOSs, dashed colored curves to the stiff limiting cases, and black/gray dotted curves to the soft limiting cases, EOS~6 and EOS~7.

\begin{figure}[!t]
    \centering
    \includegraphics[width=3.3in]{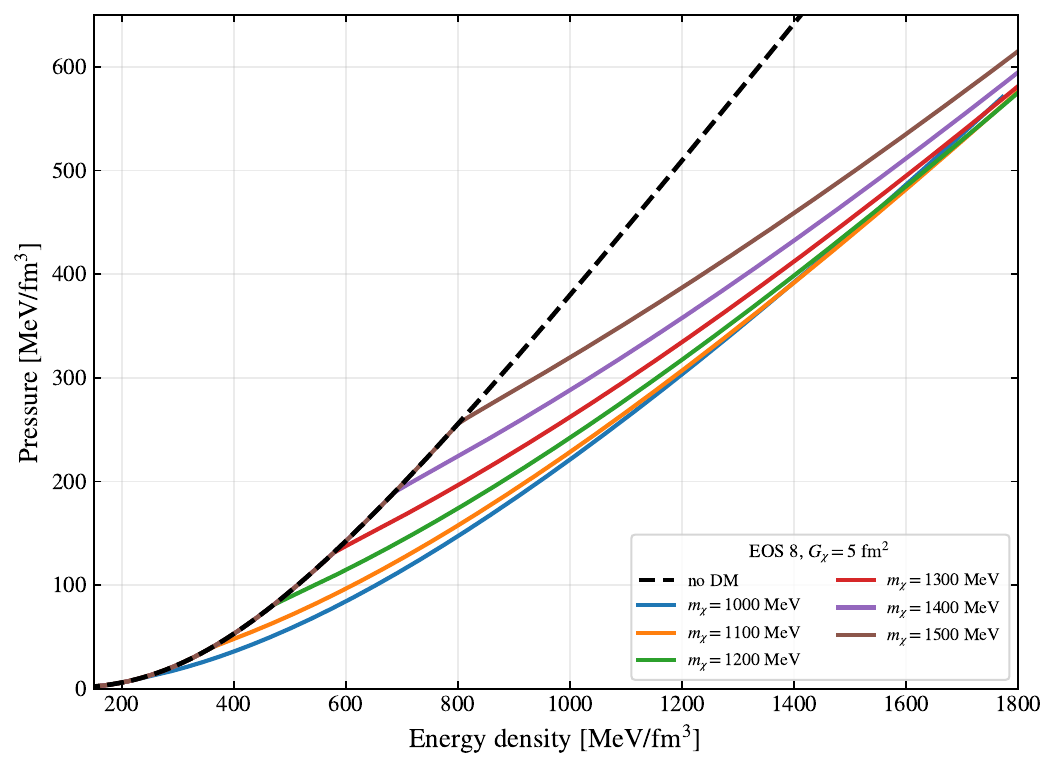}

    \vspace{0.08in}

    \includegraphics[width=3.3in]{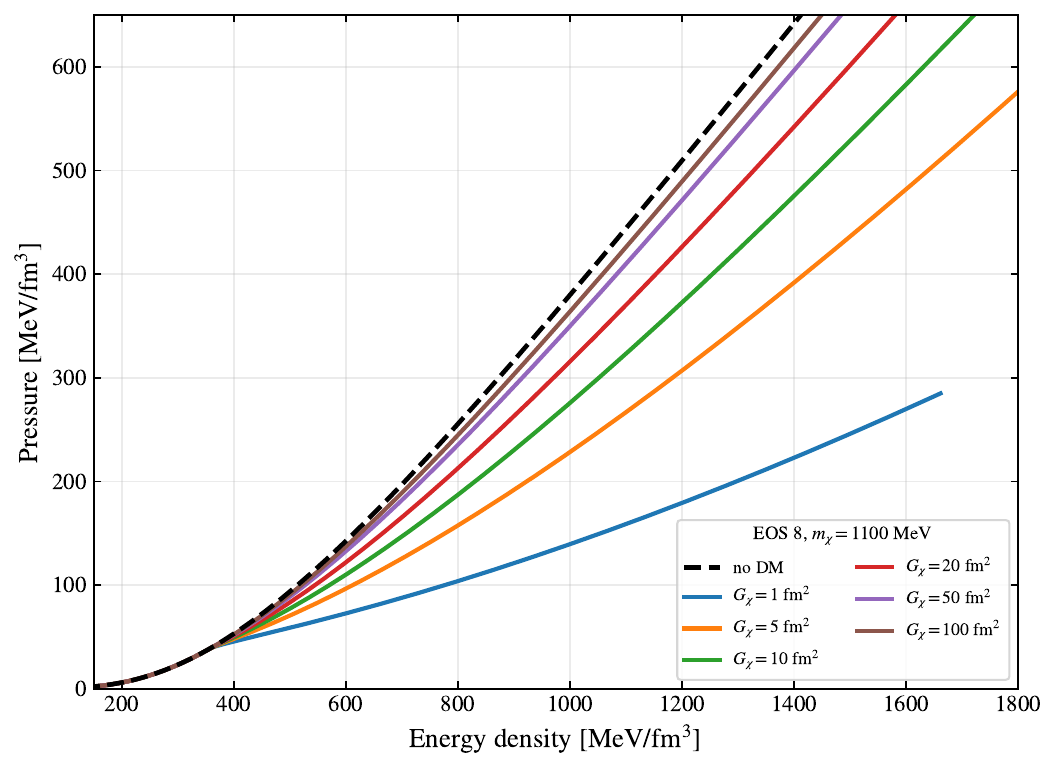}

    \vspace{0.08in}

    \includegraphics[width=3.3in]{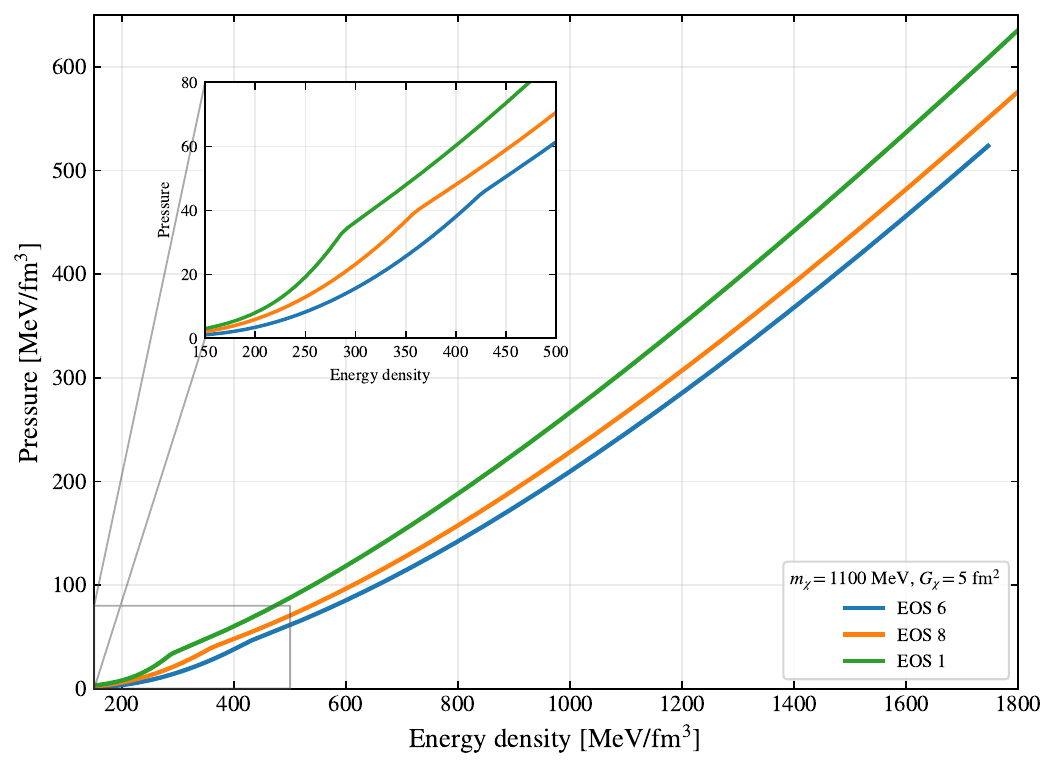}

    \caption{
Effect of neutron-to-dark-fermion conversion on the EOS. The top panel shows EOS~8 at fixed $G_\chi=5~\mathrm{fm}^{2}$ for varying $m_\chi$, while the middle panel shows EOS~8 at fixed $m_\chi=1100~\mathrm{MeV}$ for varying $G_\chi$. The black dashed curves denote the no-dark matter EOS~8 result. The bottom  panel compares EOS~6 (soft), EOS~8 (main baseline), and EOS~1 (stiff) at $m_\chi=1100~\mathrm{MeV}$ and $G_\chi=5~\mathrm{fm}^{2}$. The inset magnifies the low-energy-density region near the onset of the dark component.  Smaller $m_\chi$ and $G_\chi$ lead to stronger softening in the EOS,
while the onset and effects of conversion also depend on
the underlying EOS.
}
\label{fig:eos_dm_parameter_dependence}
\end{figure}

\section{effects of the dark Sector}
\label{sec:eos_composition_effects}

This section provides the physical background for the constraint analysis by showing how the dark component affects the dense-matter EOS, particle composition, and neutron star observables.

\subsection{EOS and Composition}

Before applying the astrophysical constraints, it is useful to illustrate how the neutron-to-dark-fermion conversion modifies the underlying EOS. 
Figure~\ref{fig:eos_dm_parameter_dependence} shows representative examples of the pressure as a function of energy density for different choices of the dark-sector parameters and for different nucleonic EOSs.

The top panel shows the effect of varying the dark-fermion mass for EOS~8 at fixed $G_\chi=5~{\rm fm}^2$. 
For smaller $m_\chi$, the conversion of neutrons into dark fermions becomes energetically more favorable, and the dark component appears more easily in dense matter. 
This reduces the pressure support relative to the no-dark matter EOS. 
As $m_\chi$ is increased, the onset of the dark component is pushed to higher density and its contribution is reduced, so the EOS moves closer to the no-dark matter limit.

The middle panel shows the complementary dependence on the dark-sector coupling at fixed $m_\chi=1100~{\rm MeV}$. 
A small value of $G_\chi$ corresponds to weak repulsion in the dark sector and therefore allows a larger dark-fermion population, producing stronger softening of the EOS. 
Increasing $G_\chi$ enhances the repulsive contribution associated with the dark component and suppresses the conversion, so the pressure increases and the EOS approaches the no-dark matter behavior. 
Thus, larger $m_\chi$ and larger $G_\chi$ both reduce the impact of the dark component, although through different physical mechanisms.

The bottom panel compares representative soft, baseline, and stiff EOSs after conversion at the common parameter point $m_\chi=1100~{\rm MeV}$ and $G_\chi=5~{\rm fm}^2$. The converted EOSs retain the pressure ordering inherited from the nucleonic sector: EOS~6 gives the lowest pressure, EOS~8 gives an intermediate behavior, and EOS~1 gives the largest pressure. The inset further shows that the change in slope marking the onset of the dark component occurs at different energy densities for the three EOSs. Thus, the qualitative stiffness ordering is preserved, while the conversion onset and the resulting EOS modification depend on the underlying nucleonic EOS.

\begin{figure}[!t]
    \centering
    \includegraphics[width=3.3in]{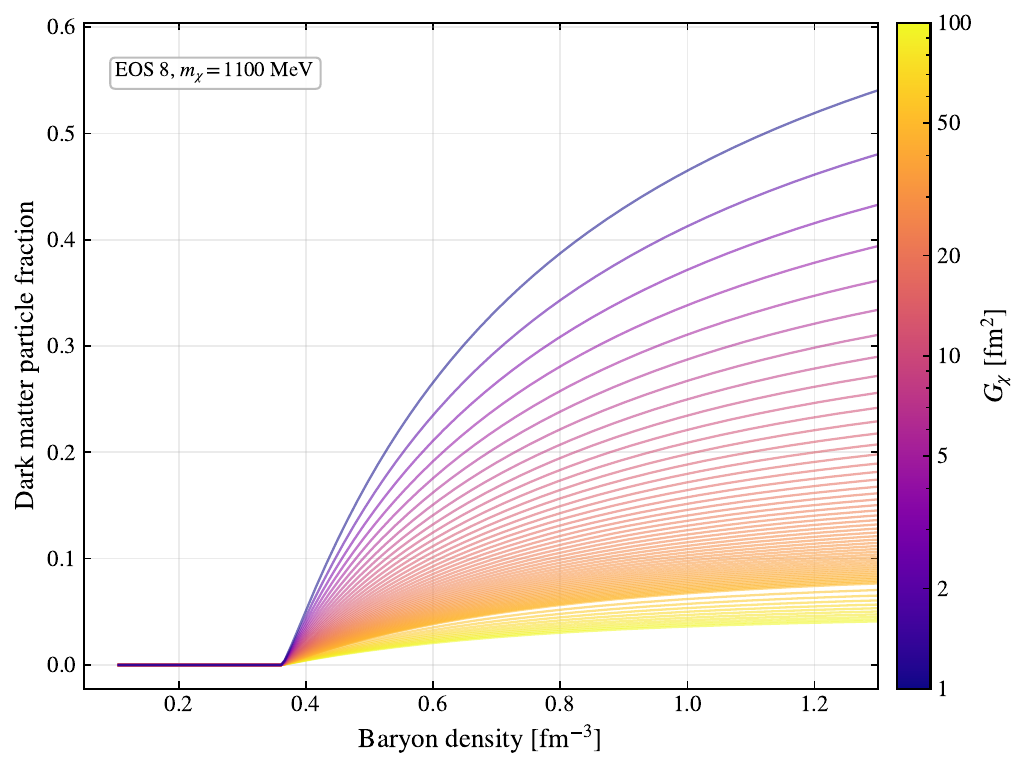}

    \vspace{0.08in}

    \includegraphics[width=3.3in]{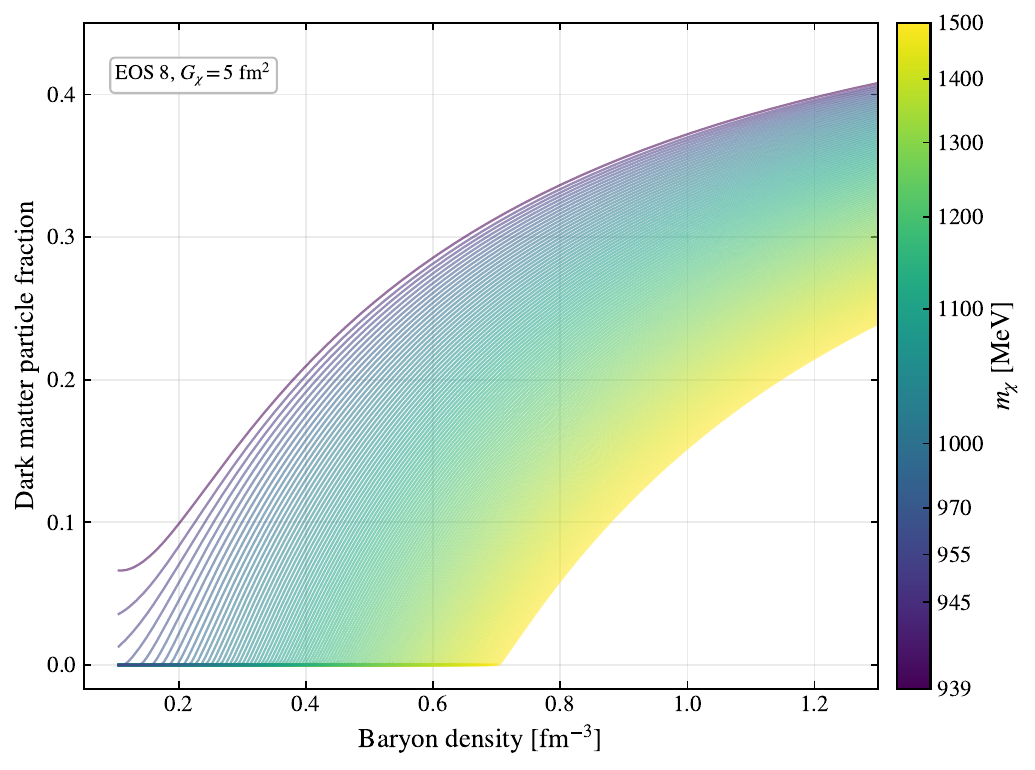}

    \vspace{0.08in}

    \includegraphics[width=3.3in]{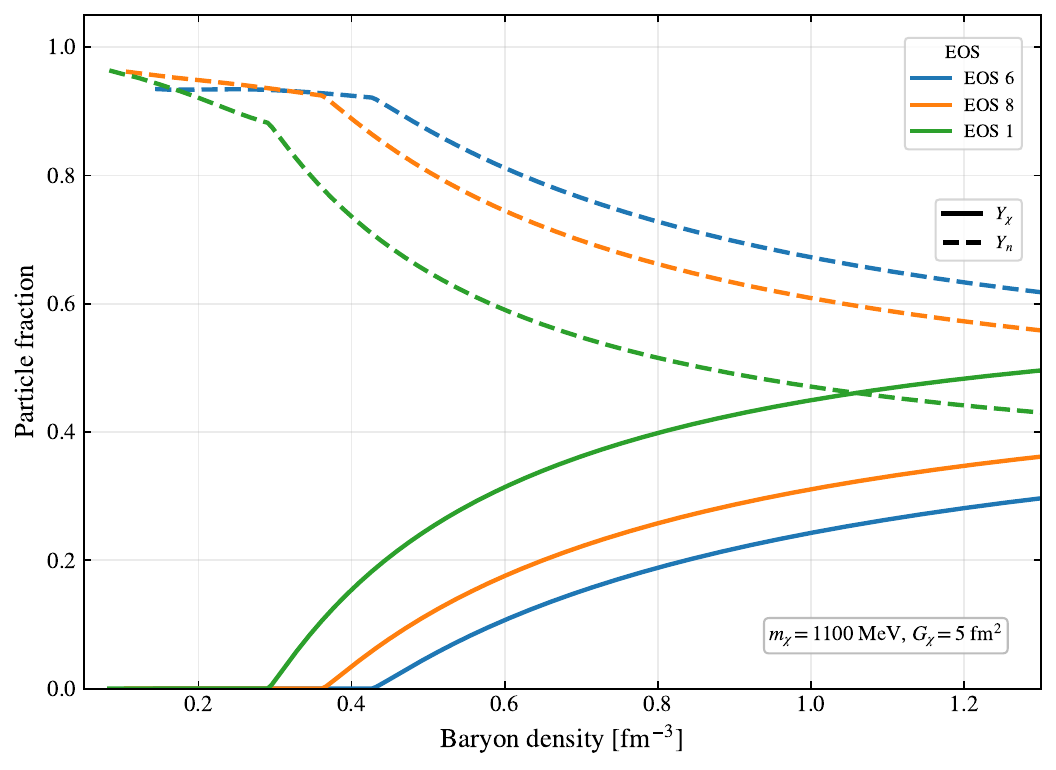}

    \caption{
    Particle fractions for the representative EOS examples. The top panel shows $Y_\chi=n_\chi/n_B$ for EOS~8 at fixed $m_\chi=1100~\mathrm{MeV}$ and varying $G_\chi$, whereas the middle panel shows EOS~8 at fixed $G_\chi=5~\mathrm{fm}^{2}$ and varying $m_\chi$. The colorbars indicate the varied parameter. The bottom panel compares EOS~6, EOS~8, and EOS~1 at $m_\chi=1100~\mathrm{MeV}$ and $G_\chi=5~\mathrm{fm}^{2}$; solid and dashed curves denote $Y_\chi$ and $Y_n$, respectively.
    }
    \label{fig:particle_fraction_examples}
\end{figure}

Figure~\ref{fig:particle_fraction_examples} illustrates how the dark-sector parameters and the underlying nucleonic EOS affect the composition of matter once neutron-to-dark-fermion conversion is allowed. 
The top and middle panels focus on EOS~8 and show the dependence of the dark-matter particle fraction \(Y_\chi=n_\chi/n_B\) on the two dark-sector parameters separately, while the bottom panel compares the behavior across representative soft, baseline, and stiff nucleonic EOSs.

The top panel shows the effect of varying the coupling \(G_\chi\) at fixed \(m_\chi=1100~{\rm MeV}\). 
For smaller \(G_\chi\), the repulsive self-interaction in the dark sector is weaker, so it is energetically easier to populate the \(\chi\) component once the conversion channel becomes available. 
As a result, the dark fraction grows more rapidly with density and reaches a larger value at high baryon density. 
By contrast, larger \(G_\chi\) suppresses the dark-sector population, so \(Y_\chi\) remains smaller throughout the density range.

The middle panel shows the complementary dependence on the dark-fermion mass \(m_\chi\) at fixed \(G_\chi=5~{\rm fm}^2\). 
Lower values of \(m_\chi\) make the conversion more favorable, so the \(\chi\) component appears at lower density and grows to a larger fraction. 
As \(m_\chi\) increases, the onset of the dark component is shifted to higher density and its abundance is reduced.

The bottom panel shows that the composition change also depends on the nucleonic EOS itself. 
At the representative point \(m_\chi=1100~{\rm MeV}\) and \(G_\chi=5~{\rm fm}^2\), the dark fraction \(Y_\chi\) is smallest for EOS~6, intermediate for EOS~8, and largest for EOS~1, while the neutron fraction \(Y_n\) decreases in the opposite order. 
In other words, EOSs that favor a larger dark-sector population exhibit a correspondingly stronger depletion of the neutron component. 
This comparison makes clear that the same dark-sector parameters do not lead to identical matter composition across different nucleonic inputs.

\subsection{Illustrative effect of dark conversion on stellar observables}
\label{subsec:illustrative_dm_effects}

\begin{figure*}[!t]
    \centering

    \begin{minipage}{0.48\textwidth}
        \centering
        \includegraphics[width=\textwidth]{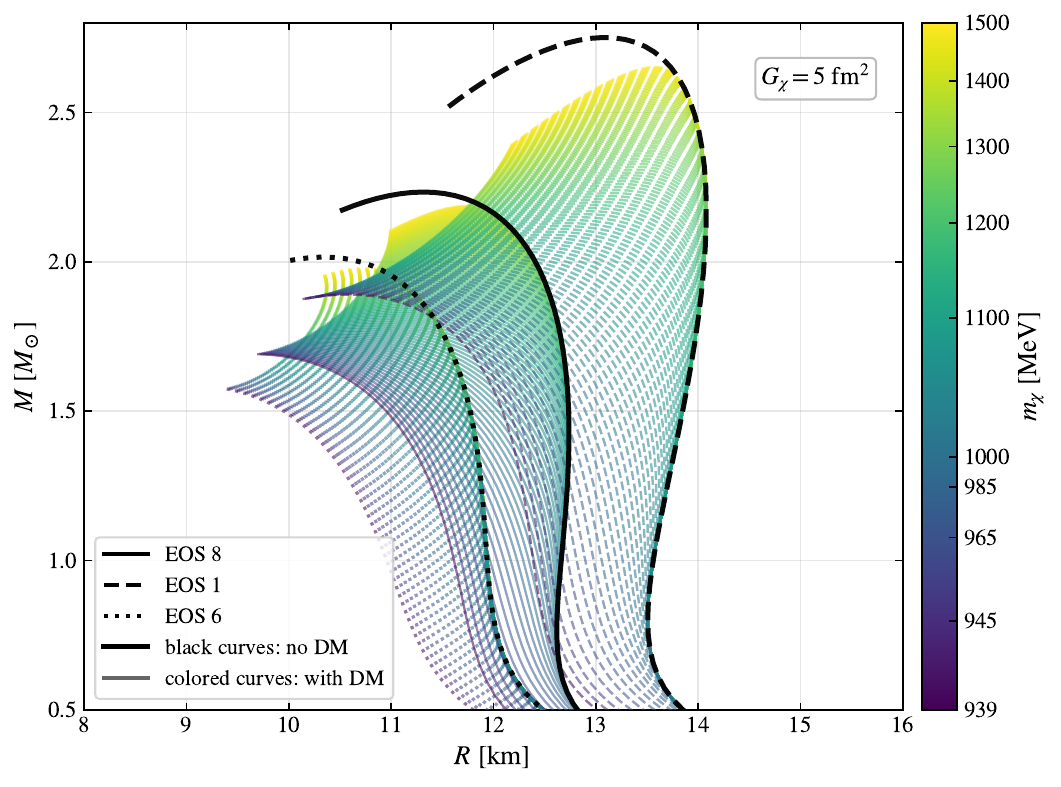}
        \vspace{-0.08in}
        \centerline{\small (a)}
    \end{minipage}
    \hfill
    \begin{minipage}{0.48\textwidth}
        \centering
        \includegraphics[width=\textwidth]{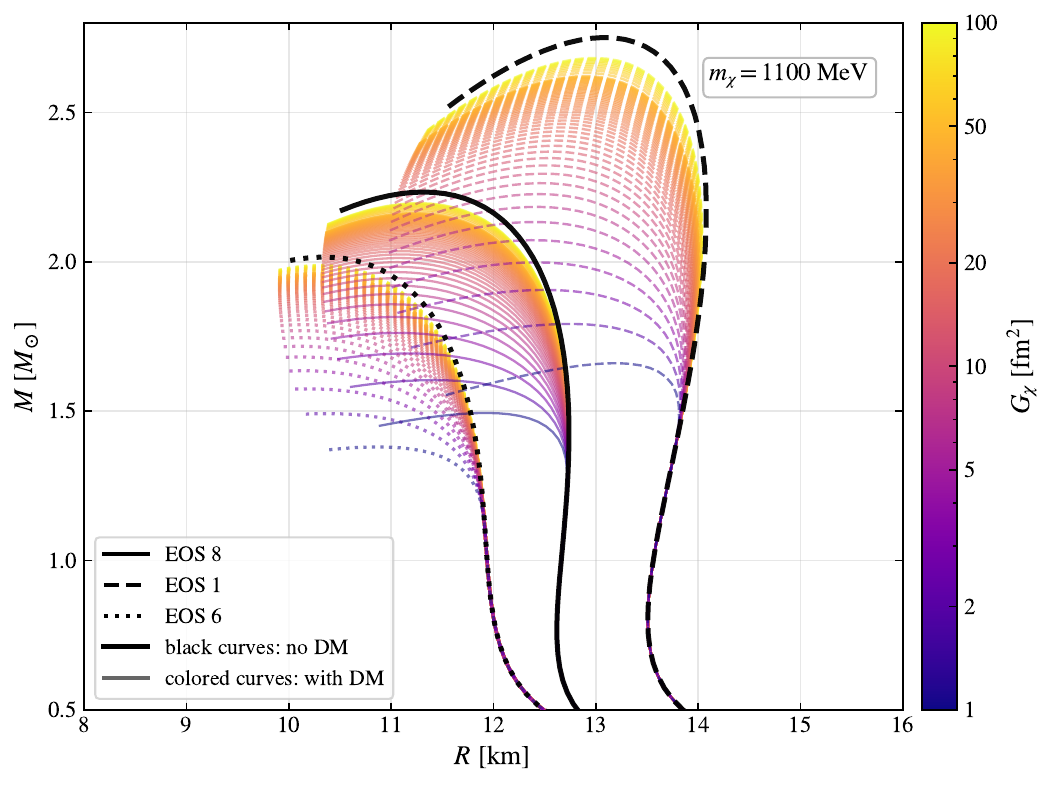}
        \vspace{-0.08in}
        \centerline{\small (b)}
    \end{minipage}

    \vspace{0.12in}

    \begin{minipage}{0.48\textwidth}
        \centering
        \includegraphics[width=\textwidth]{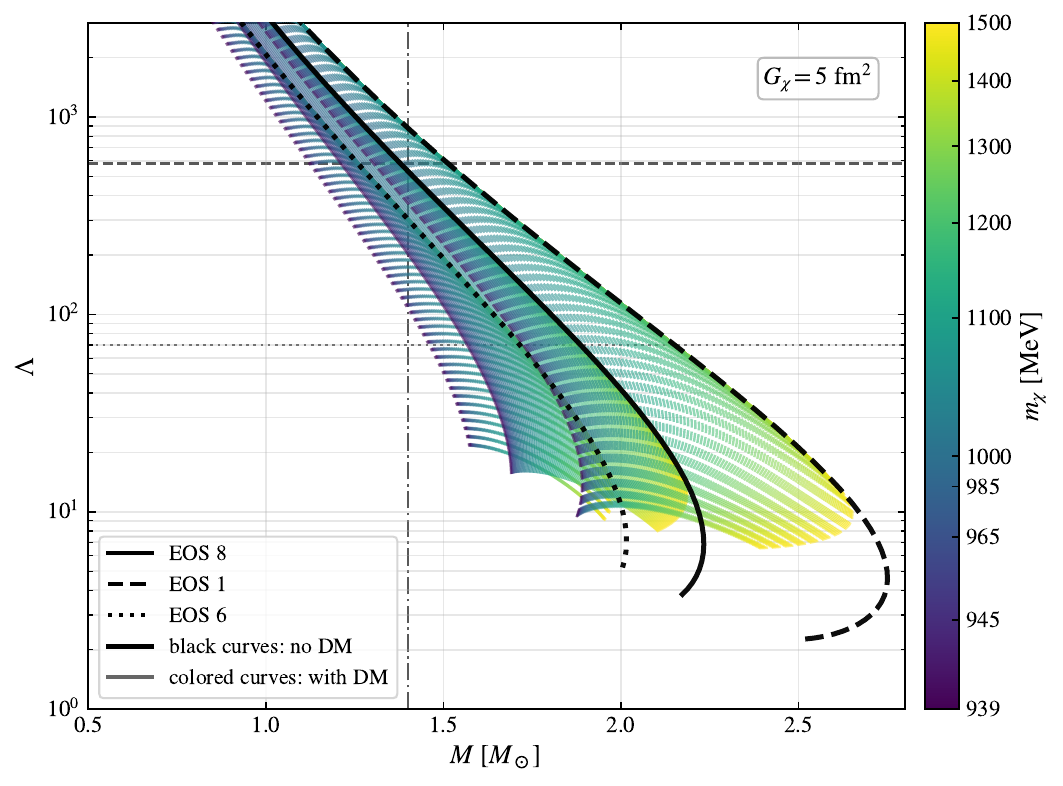}
        \vspace{-0.08in}
        \centerline{\small (c)}
    \end{minipage}
    \hfill
    \begin{minipage}{0.48\textwidth}
        \centering
        \includegraphics[width=\textwidth]{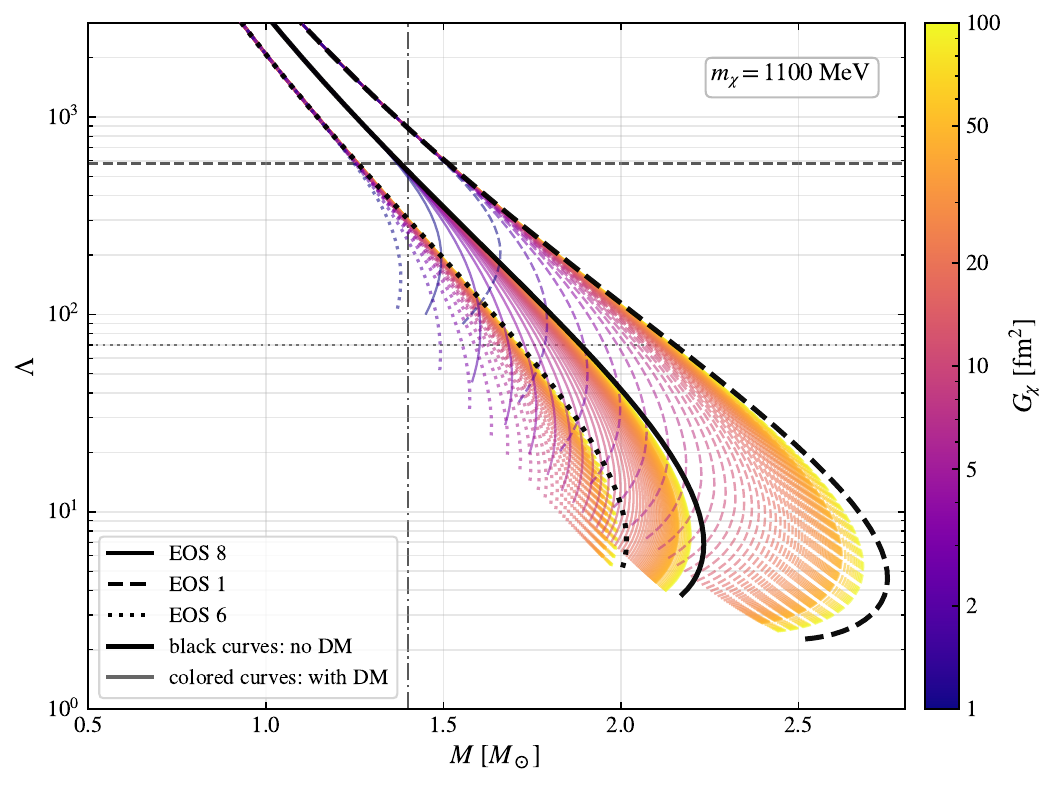}
        \vspace{-0.08in}
        \centerline{\small (d)}
    \end{minipage}

    \caption{
   Mass--radius relations (upper row) and dimensionless tidal deformability as a function of mass (lower row) for EOS~8 (solid), EOS~1 (dashed), and EOS~6 (dotted). Panels (a) and (c) show the $m_\chi$ scan at fixed $G_\chi=5~\mathrm{fm}^{2}$, whereas panels (b) and (d) show the $G_\chi$ scan at fixed $m_\chi=1100~\mathrm{MeV}$. The colorbars indicate the varied parameter. Colored and black curves denote the converted and no-dark matter sequences, respectively. In the lower panels, the vertical line marks $M=1.4\,M_\odot$, and the horizontal lines show the adopted tidal-deformability bounds.
    }
    \label{fig:mr_lambdam_parameter_dependence}
\end{figure*}

  Having examined how neutron-to-dark-fermion conversion
affects the EOS and particle composition, we now turn
to its impact on stellar observables for representative EOSs.  Figure~\ref{fig:mr_lambdam_parameter_dependence} presents the stellar sequences obtained with
dark-fermion conversion alongside their corresponding
no-dark matter sequences.  The purpose of this figure is not yet to apply the full exclusion analysis, but to illustrate how the microscopic dark-sector parameters \(m_\chi\) and \(G_\chi\) affect the mass--radius and \(\Lambda\)--\(M\) relations before the EOS-by-EOS constraint analysis in Secs.~\ref{sec:main_baseline_results}--\ref{sec:stiff_results}.

Figure~\ref{fig:mr_lambdam_parameter_dependence} compares three representative nucleonic inputs: EOS~6 as a soft limiting case, EOS~8 as a main-baseline reference case, and EOS~1 as a stiff limiting case.
For each EOS, the black curve shows the corresponding no-dark matter stellar sequence, while the colored curves show the sequences obtained after imposing neutron-to-dark-fermion conversion for the scanned dark-sector parameters. Compared with the EOS and composition shown in Figs. ~\ref{fig:eos_dm_parameter_dependence} and \ref{fig:particle_fraction_examples}, the present figure demonstrates how those microscopic changes propagate to directly observable neutron star properties.

The figures to the left show the effect of varying the dark-fermion mass \(m_\chi\) at fixed coupling \(G_\chi=5~{\rm fm}^2\). 
As \(m_\chi\) increases, the onset and abundance of the dark component are suppressed, and the corresponding mass--radius and \(\Lambda\)--\(M\) curves move closer to the no-dark matter reference sequence.

The figures to the right show  the complementary dependence on the dark-sector coupling \(G_\chi\) at fixed \(m_\chi=1100~{\rm MeV}\).  
Increasing \(G_\chi\) suppresses the dark fermion fraction, so the stellar sequences progressively approach the no-dark matter limit.

The lower panels show that the tidal deformability is especially sensitive to these changes. 
Since \(\Lambda\) depends strongly on stellar compactness, even moderate shifts in radius and internal structure can produce a visible displacement of the \(\Lambda\)--\(M\) curves. 
The guide line at \(M=1.4M_\odot\) indicates where the canonical tidal deformability \(\Lambda_{1.4}\) is evaluated, while the horizontal lines show the adopted tidal-deformability interval. 
These panels therefore make clear why \(\Lambda_{1.4}\) provides an important constraint on the dark-sector parameter space, in addition to the maximum-mass requirement.

The comparison among EOS~6, EOS~8, and EOS~1 reveals a common qualitative response to neutron--dark-fermion conversion: the converted mass--radius and tidal-deformability sequences preserve the soft-to-stiff ordering of their no-dark matter counterparts. However, the response is not identical across the EOSs, because both the conversion onset and the equilibrium dark-fermion fraction depend on the nucleonic background. Consequently, the same values of $(m_\chi,G_\chi)$ can produce different degrees of modification in the stellar observables, even though the general EOS ordering remains unchanged.

Overall, increases in $m_\chi$ and $G_\chi$ both drive the system toward the purely nucleonic limit, again through different mechanisms: increasing $m_\chi$ primarily delays the onset of conversion, whereas increasing $G_\chi$ suppresses the equilibrium dark-particle abundance after conversion becomes possible.

\section{Constraint strategy and notation}
\label{sec:constraint_strategy}

We now apply the single-$\chi$ conversion model, defined by $\mu_\chi=\mu_n$, over the full scanned range considered in this work, $m_\chi=939$--$1500~\mathrm{MeV}$ and $G_\chi=1$--$100~{\rm fm}^2$.   The purpose of this systematic scan is to identify
which combinations of dark-sector parameters are
excluded by multimessenger observations and to
assess how these exclusions depend on the underlying
nucleonic EOS.  For each nucleonic EOS and each grid point in $(m_\chi,G_\chi)$, the corresponding stellar sequence is tested against the massive-star requirement \cite{Fonseca:2021wxt}, the tidal deformability limits \cite{LIGOScientific:2018cki}, and, separately, the Illinois--Maryland and X-PSI/Amsterdam NICER mass--radius inferences \cite{Miller:2026aju}.

Throughout the following analysis, we refer to
$$
M_{\max}\geq 2M_\odot,
\qquad
70\leq \Lambda_{1.4}\leq 580
$$
as the \emph{basic} constraint set. We then apply the two independent NICER inference sets separately and on equal footing. For the exclusion analysis, we add all four
Illinois--Maryland 95\% mass--radius regions to the
basic constraints and, independently, repeat the same
procedure using all four X-PSI/Amsterdam 95\% regions.
We refer to the combination of the basic constraints
with all four 95\% regions as the conservative
basic+all-95\% exclusion criterion for each NICER
inference set. This criterion is used to define the
excluded and not-excluded regions for the corresponding
NICER analysis. The term conservative reflects our use
of the broader 95\% credible regions rather than the
more restrictive 68\% regions to define exclusions.
In particular, a stellar sequence should not be excluded
on the basis of a NICER mass--radius inference simply
because it does not intersect the corresponding 68\%
credible region, provided that it still intersects the
95\% region.

In addition to the conservative analyses, we perform stricter consistency tests only for the 13 main-baseline EOSs, which constitute our primary reference set. These tests identify the parts of the conservatively not-excluded parameter space that remain compatible with the tighter credible regions of each NICER inference set and illustrate how the dark sector shifts the corresponding mass--radius sequences. For the Illinois--Maryland analysis, this test combines the basic constraints with the 68\% regions of PSR J0030+0451, PSR J0740+6620, PSR J0437--4715, and PSR J0614--3329. For the X-PSI/Amsterdam analysis, it combines the basic constraints with the 68\% regions of PSR J0030+0451, PSR J0740+6620, and PSR J0437--4715, while retaining the 95\% region for the compact source PSR J0614--3329.  This is because its 68\%
credible region is very compact, and imposing this
more restrictive region together with the other
requirements leaves no consistent parameter points
in the scanned domain.  These stricter tests are used only as consistency checks and not as exclusion criteria. For the soft and stiff EOS sets, which are included as limiting cases to examine how the conservative excluded and not-excluded regions depend on the nucleonic EOS, we do not present a detailed $68\%$ consistency analysis and instead provide only brief comments on their $68\%$ behavior.

For each EOS group and each conservative NICER analysis, $N_{\rm excl}$ denotes the number of EOSs for which the corresponding part of parameter space is excluded in that analysis. Thus, $N_{\rm excl}=0$ defines the common not-excluded region in the  $(m_\chi,G_\chi)$ plane, across all EOSs in that group. This region should be interpreted as parameter space not excluded by the present constraints, rather than as uniquely allowed in an absolute sense. Note that the terms ``common'' and ``overlapped'' refer to regions shared by all EOSs within the specified group and are restricted to this RMF ensemble; they do not imply model independence beyond the adopted nucleonic framework. For the stricter consistency tests applied to the main-baseline set, $N_{\rm cons}$ denotes the number of EOSs whose stellar sequences remain consistent with the corresponding requirements.

In the following sections, we first present the Illinois--Maryland and X-PSI/Amsterdam conservative analyses separately for the main-baseline, soft, and stiff EOS groups. The corresponding stricter consistency tests are then presented only for the main-baseline set. We finally compare the EOS-overlapped conservative exclusions obtained from the two NICER inference sets for the complete 21-EOS ensemble.

\section{Main baseline Equations of state}
\label{sec:main_baseline_results}

We first consider the 13 main baseline EOSs, which form the reference set for the dark-sector constraints.  Here we focus on how the multimessenger constraints reshape the dark matter parameter space once neutron-to-dark-fermion conversion is included. The Illinois--Maryland and X-PSI/Amsterdam analyses are presented separately in the following subsections, followed by the corresponding stricter consistency tests.

\subsection{Illinois–Maryland analysis}

\begin{figure}[t]
\centering
\includegraphics[width=\linewidth]{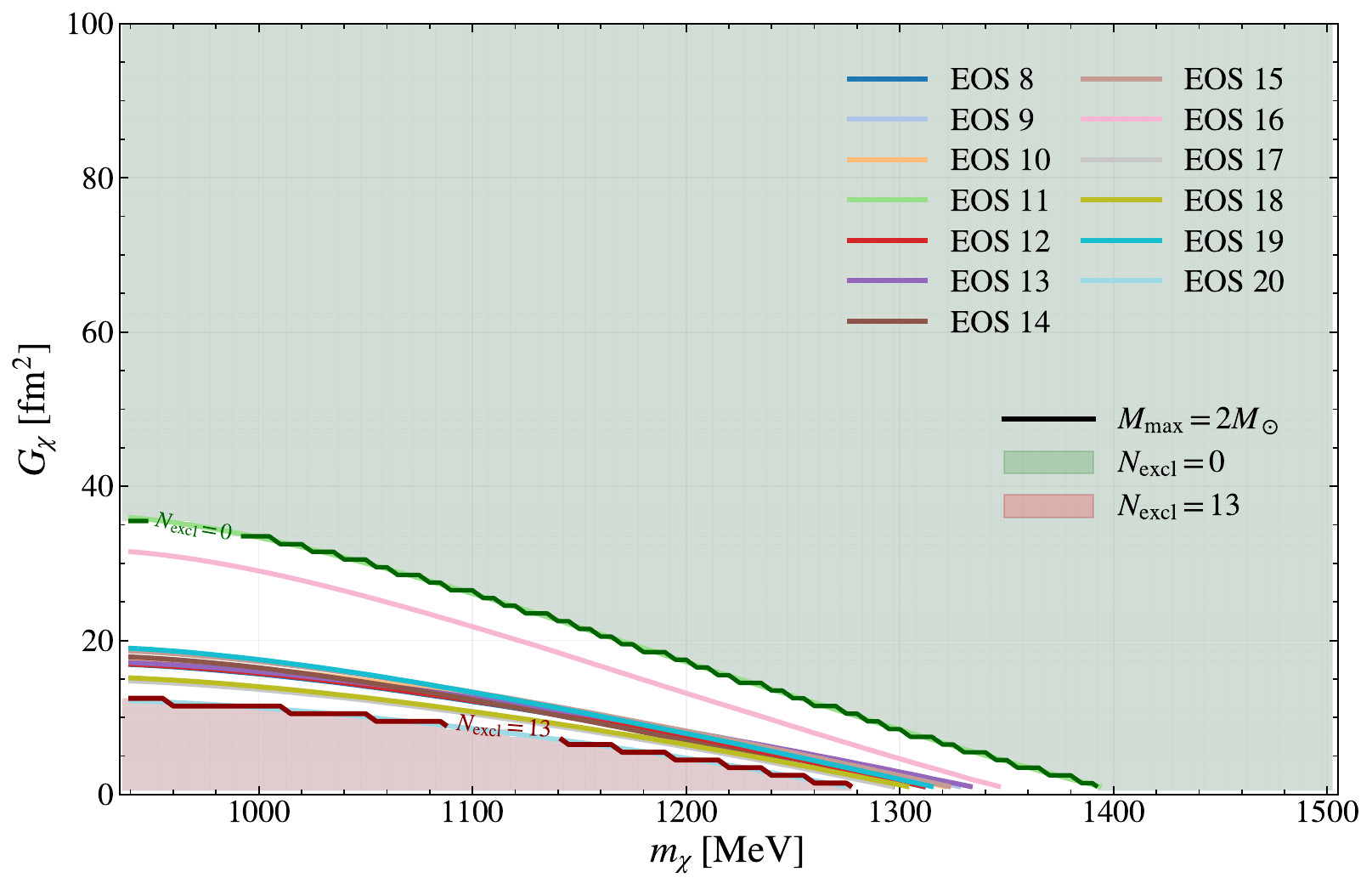}
\caption{
Exclusion boundaries in the $(m_\chi,G_\chi)$ plane
for the 13 main baseline EOSs in the conservative
basic+all-95\% Illinois--Maryland analysis.
The tidal-deformability constraint and all four
Illinois--Maryland 95\% mass--radius regions introduce
no additional exclusions beyond the $M_{\max}$
constraint. Therefore, the solid curves show both
the $M_{\max}=2M_\odot$ boundaries and the final
conservative Illinois--Maryland boundaries.
The green region represents the common
not-excluded region ($N_{\rm excl}=0$), while the
red-shaded region represents the common excluded
region ($N_{\rm excl}=13$). The unshaded (white)
region corresponds to $0<N_{\rm excl}<13$, where
exclusion depends on the nucleonic EOS.
}
\label{fig:illinois_maryland_boundaries_baseline13}
\end{figure}

We first apply the conservative basic+all-95\% criterion using the four Illinois--Maryland NICER mass--radius inferences. For all 13 main-baseline EOSs, the tidal-deformability requirement removes no parameter points beyond those already excluded by the $M_{\max}\geq 2M_\odot$ condition. Moreover, none of the four Illinois--Maryland 95\% regions introduces any additional exclusion among the remaining points. The conservative Illinois--Maryland result therefore
coincides exactly with that obtained from the
maximum-mass constraint alone, as shown in Fig.~\ref{fig:illinois_maryland_boundaries_baseline13}.

The solid $M_{\max}=2M_\odot$ curves mark the boundary
of the requirement that the maximum neutron-star mass
be at least two solar masses. These curves therefore also
represent the final conservative Illinois--Maryland
exclusion boundaries.  The green region, labeled $N_{\rm excl}=0$,
represents the common not-excluded region, while
the red-shaded region at small $m_\chi$ and weak
$G_\chi$, labeled $N_{\rm excl}=13$, is excluded
by all 13 EOSs. The green and red curves mark
the boundaries of these two regions, respectively. The unshaded region between the green and red areas is EOS dependent, with $0<N_{\rm excl}<13$: for each colored EOS boundary, parameter points below the curve are excluded for that EOS, whereas points above it are not excluded under the conservative Illinois--Maryland analysis.
Across the full scanned mass range, all points with $G_\chi\geq36~{\rm fm}^2$ belong to the not-excluded region common to all 13 EOSs.  Similarly, from $m_\chi=1395~\mathrm{MeV}$ upward, all scanned values of $G_\chi$ are in the common not-excluded region. The common-excluded region reaches $G_\chi=12~{\rm fm}^2$ for $m_\chi=939$--$955~\mathrm{MeV}$ and gradually narrows to $G_\chi=1~{\rm fm}^2$ at $m_\chi=1275~\mathrm{MeV}$.

\subsection{X-PSI/Amsterdam analysis}

\begin{figure}[t]
\centering
\includegraphics[width=\linewidth]{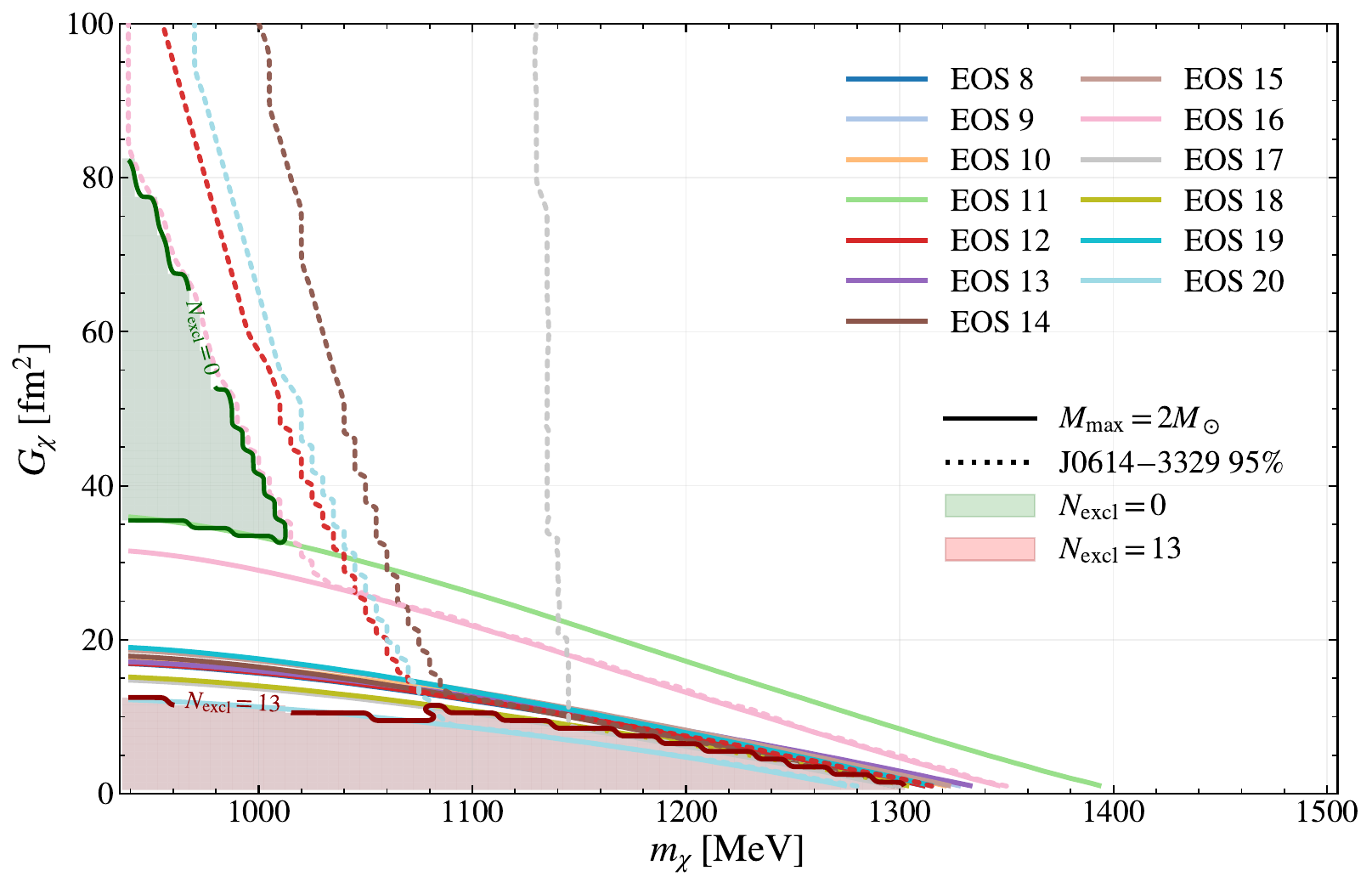}
\caption{
Exclusion boundaries in the $(m_\chi,G_\chi)$ plane for the 13 main baseline EOSs under the conservative basic+all-95\% X-PSI/Amsterdam analysis. Solid curves show the $M_{\max}=2M_\odot$ boundaries, dotted curves show the active PSR J0614--3329 95\% boundary, and the green/red regions correspond to $N_{\rm excl}=0$/$13$.
}
\label{fig:amsterdam_boundaries_baseline13}
\end{figure}

We next apply the same conservative basic+all-95\% criterion using the four X-PSI/Amsterdam NICER mass--radius inferences. In contrast to the Illinois--Maryland result, the X-PSI/Amsterdam regions introduce additional exclusions in part of the scanned parameter space. Figure~\ref{fig:amsterdam_boundaries_baseline13} shows the resulting conservative exclusion structure for the 13 main baseline EOSs. The visible limiting boundaries are mainly controlled by the maximum-mass condition and the 95\% mass--radius region of PSR J0614--3329. The low-$G_\chi$ region is excluded because weak dark self-repulsion cannot compensate for the EOS softening caused by conversion. For EOS~12, EOS~14, EOS~16, EOS~17, and EOS~20, the compact J0614--3329 region also becomes active. For these EOSs, the dark component must make the stellar sequence sufficiently compact to enter the 95\% credible region without reducing $M_{\max}$ below $2M_\odot$. The tidal-deformability bound and the other X-PSI/Amsterdam 95\% regions are included in the classification, but they do not define the visible limiting boundaries for this EOS group.

Figure~\ref{fig:amsterdam_boundaries_baseline13} displays the individual limiting boundaries. The solid curves denote the $M_{\max}=2M_\odot$ boundaries
of the $M_{\max}\geq 2M_\odot$ requirement; for each EOS, the region below the corresponding solid curve is excluded because the dark-matter-admixed stellar sequence cannot support a $2M_\odot$ neutron star. The dotted curves denote the active 95\% NICER boundary from PSR J0614--3329. For the EOSs with dotted boundaries, and among the points satisfying the maximum-mass condition, the region to the left of the corresponding dotted curve is compatible with the J0614--3329 95\% mass--radius constraint, whereas the region to its right and upper right is excluded by this compact-source constraint.  Thus, in these cases, although some dark-sector parameter points are
excluded by the observational constraints,
neutron-to-dark-fermion conversion can also improve
the compatibility of the nucleonic stellar sequences
that initially lie outside the 95\% credible region
of PSR J0614--3329 by shifting the mass--radius
sequences toward smaller radii.

The shaded regions in Fig.~\ref{fig:amsterdam_boundaries_baseline13} summarize the overlap across the 13 EOSs. The green region corresponds to $N_{\rm excl}=0$, meaning that none of the main baseline EOSs is excluded at those parameter points. The red region corresponds to $N_{\rm excl}=13$, meaning that all 13 main baseline EOSs are excluded. The unshaded region between them is EOS dependent, with $0<N_{\rm excl}<13$: some EOSs are excluded at a given parameter point, while others are not. The red region therefore represents the EOS-common excluded region, where changing the nucleonic EOS within the main baseline set cannot recover compatibility.

As discussed in Sec.~\ref{sec:eos_composition_effects}, the excluded and not-excluded regions in Fig.~\ref{fig:amsterdam_boundaries_baseline13} arise from the competing effects of $m_\chi$ and $G_\chi$. The common not-excluded region is determined by a balance: $G_\chi$ must be sufficiently large to preserve $M_{\max}\geq2M_\odot$, but, for the EOSs constrained by PSR J0614--3329, not so large that it suppresses the dark component and prevents the stellar sequence from entering the corresponding 95\% credible region.

Quantitatively, the EOS-common excluded region, $N_{\rm excl}=13$, has the minimum--maximum extents
$$
m_\chi=939\text{--}1300~{\rm MeV},
\qquad
G_\chi=1\text{--}12~{\rm fm}^2 .
$$
The common not-excluded region, $N_{\rm excl}=0$, has the minimum--maximum extents
$$
m_\chi=939\text{--}1010~{\rm MeV},
\qquad
G_\chi=33\text{--}80~{\rm fm}^2 .
$$
These intervals specify only the minimum and maximum
values of $m_\chi$ and $G_\chi$ among the grid points
belonging to each region. They do not define rectangular
regions containing every combination of $m_\chi$ and
$G_\chi$ within these limits, since the actual excluded
and not-excluded regions have the boundaries shown
in Fig.~\ref{fig:amsterdam_boundaries_baseline13}.

Comparing Figs.~\ref{fig:illinois_maryland_boundaries_baseline13}
and \ref{fig:amsterdam_boundaries_baseline13}, the
Illinois--Maryland analysis leaves a broad common
not-excluded region, since the maximum-mass requirement
determines the exclusion boundaries. In contrast, the
compact PSR J0614--3329 constraint in the X-PSI/Amsterdam
analysis excludes parameter points for which the
dark-sector effects are too weak to make the stellar
sequences of some nucleonic EOSs sufficiently compact.
As a result, the common not-excluded region becomes
much smaller and is restricted to low $m_\chi$ and
intermediate $G_\chi$, while the common excluded regions
at low $G_\chi$ remain similar in both analyses.

\subsection{Stricter consistency tests}

We now turn to the stricter consistency tests, which
examine whether parameter points remaining after the
conservative analysis also satisfy tighter NICER
mass--radius constraints. These tests are used to
assess consistency rather than to define additional
exclusions. We first consider the
Illinois--Maryland analysis, replacing the four 95\%
credible regions with their corresponding 68\% regions.  At the 68\% level, all parameter points that remain after the conservative Illinois--Maryland analysis also remain consistent with the four tighter mass--radius regions for each of the 13 main baseline EOSs. The stricter Illinois--Maryland test therefore produces no additional reduction of the parameter space. Its strict-consistency region coincides with the corresponding conservative not-excluded region shown in Fig.~\ref{fig:illinois_maryland_boundaries_baseline13}, and no separate strict-consistency parameter-space plot is required.

\begin{figure*}[!t]
\centering
\includegraphics[width=\textwidth]{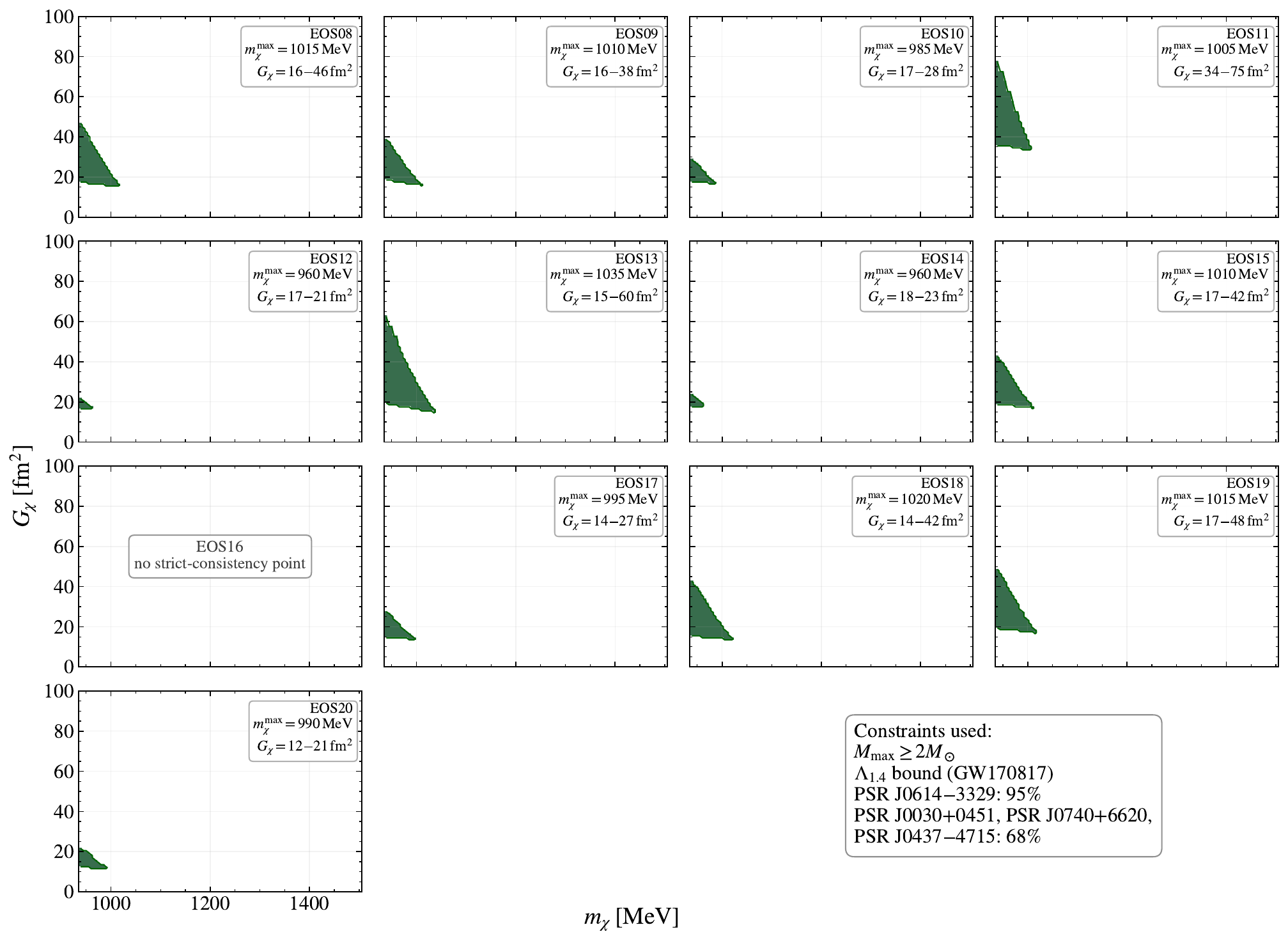}
\caption{
EOS-by-EOS strict-consistency regions for the 13 main baseline EOSs under the strict X-PSI/Amsterdam criterion. Green points satisfy the basic constraints, the 68\% regions of PSR J0030+0451, PSR J0740+6620, and PSR J0437--4715, and the 95\% region of PSR J0614--3329. Each panel shows the largest consistent $m_\chi$ and the corresponding $G_\chi$ interval.
}
\label{fig:eos_by_eos_consistent_regions}
\end{figure*}

The stricter X-PSI/Amsterdam consistency test gives a narrower and more EOS-dependent picture, as illustrated in Fig.~\ref{fig:eos_by_eos_consistent_regions}. This test is not a conservative exclusion criterion; rather, it identifies the part of the parameter space that satisfies the stricter X-PSI/Amsterdam mass--radius consistency requirements. Figure~\ref{fig:eos_by_eos_consistent_regions} displays the resulting strict-consistency region separately for each EOS, showing directly how this region varies among the EOSs. EOS~13 has the largest such region, extending up to $m_\chi^{\max}=1035~{\rm MeV}$. EOS~8, EOS~18, and EOS~19 also retain sizeable regions. In contrast, EOS~12 and EOS~14 retain only much smaller domains, while EOS~16 has no point consistent with this strict criterion.

A common feature of all non-empty panels in Fig.~\ref{fig:eos_by_eos_consistent_regions} is that the masses in the strict-consistency regions lie close to the neutron mass.  Although the scan extends up to $m_\chi=1500~{\rm MeV}$,  the regions satisfying the strict X-PSI/Amsterdam
consistency requirements are confined to much lower masses.  The corresponding values of $G_\chi$ are also restricted to finite intermediate intervals.

\begin{figure}[t]
\centering
\includegraphics[width=\columnwidth]{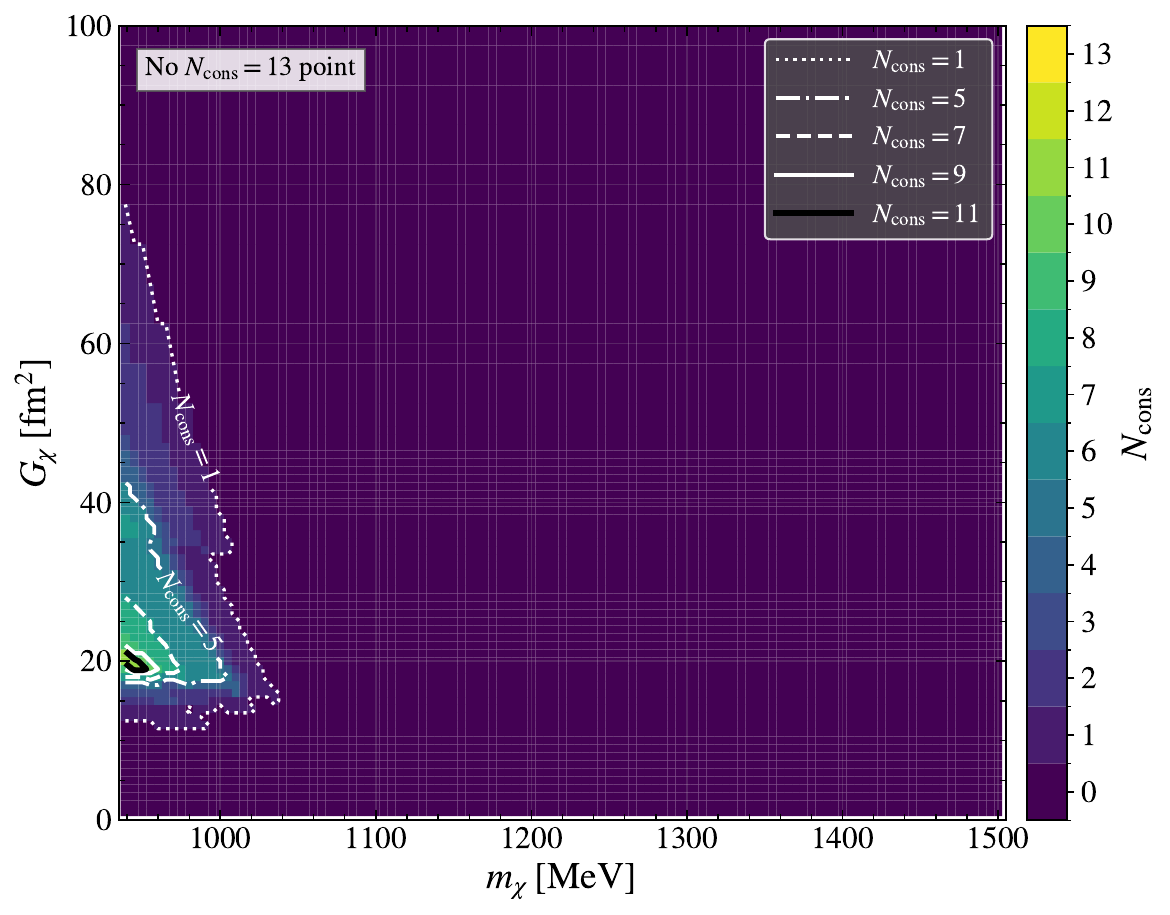}
\caption{
EOS-overlap strict-consistency map for the 13 main baseline EOSs. The color scale gives $N_{\rm cons}$, the number of baseline EOSs satisfying the strict X-PSI/Amsterdam criterion at each point. No point has $N_{\rm cons}=13$; the $N_{\rm cons}=11$ core is confined to $m_\chi=939$--$950~{\rm MeV}$ and $G_\chi=19$--$21~{\rm fm}^2$.
}
\label{fig:robust_consistency_heatmap_amsterdam}
\end{figure}

Figure~\ref{fig:robust_consistency_heatmap_amsterdam} depicts the baseline EOSs that remain consistent under the strict X-PSI/Amsterdam test.  No part of the scanned parameter space is consistent with all 13 baseline EOSs, but a small region remains consistent for 11 of them. This $N_{\rm cons}=11$ region is confined to
$$
m_\chi=939\text{--}950~{\rm MeV},
\qquad
G_\chi=19\text{--}21~{\rm fm}^2 .
$$
Thus, the region compatible with the largest number of baseline EOSs under the strict X-PSI/Amsterdam consistency test lies close to the neutron-mass threshold and requires an intermediate dark self-repulsion. This should be read as a consistency statement, not as a conservative exclusion.

\begin{figure*}[!t]
\centering
\begin{minipage}{0.49\textwidth}
\centering
\includegraphics[width=\linewidth]{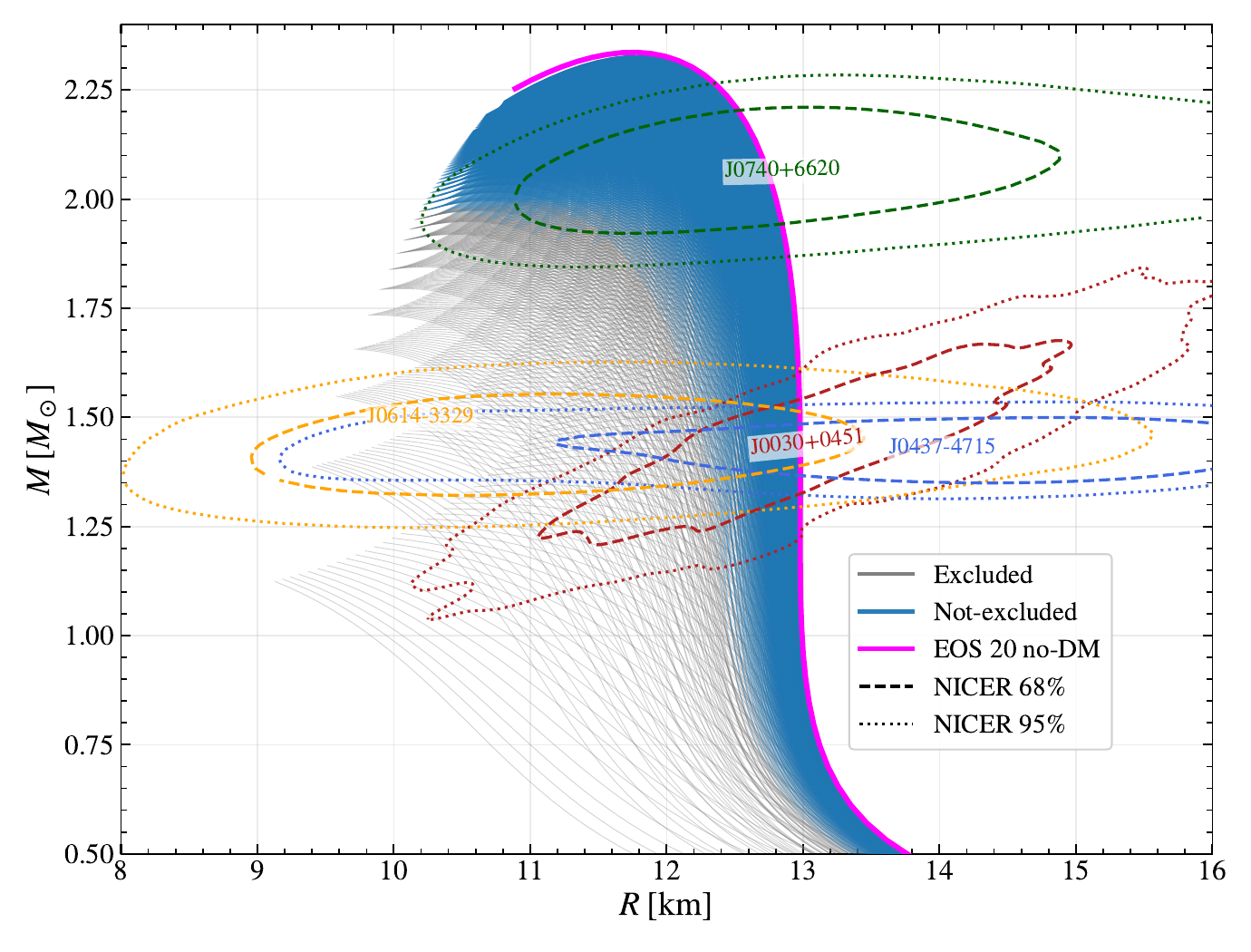}
\par\smallskip
{\small (a) Illinois--Maryland}
\end{minipage}
\hfill
\begin{minipage}{0.49\textwidth}
\centering
\includegraphics[width=\linewidth]{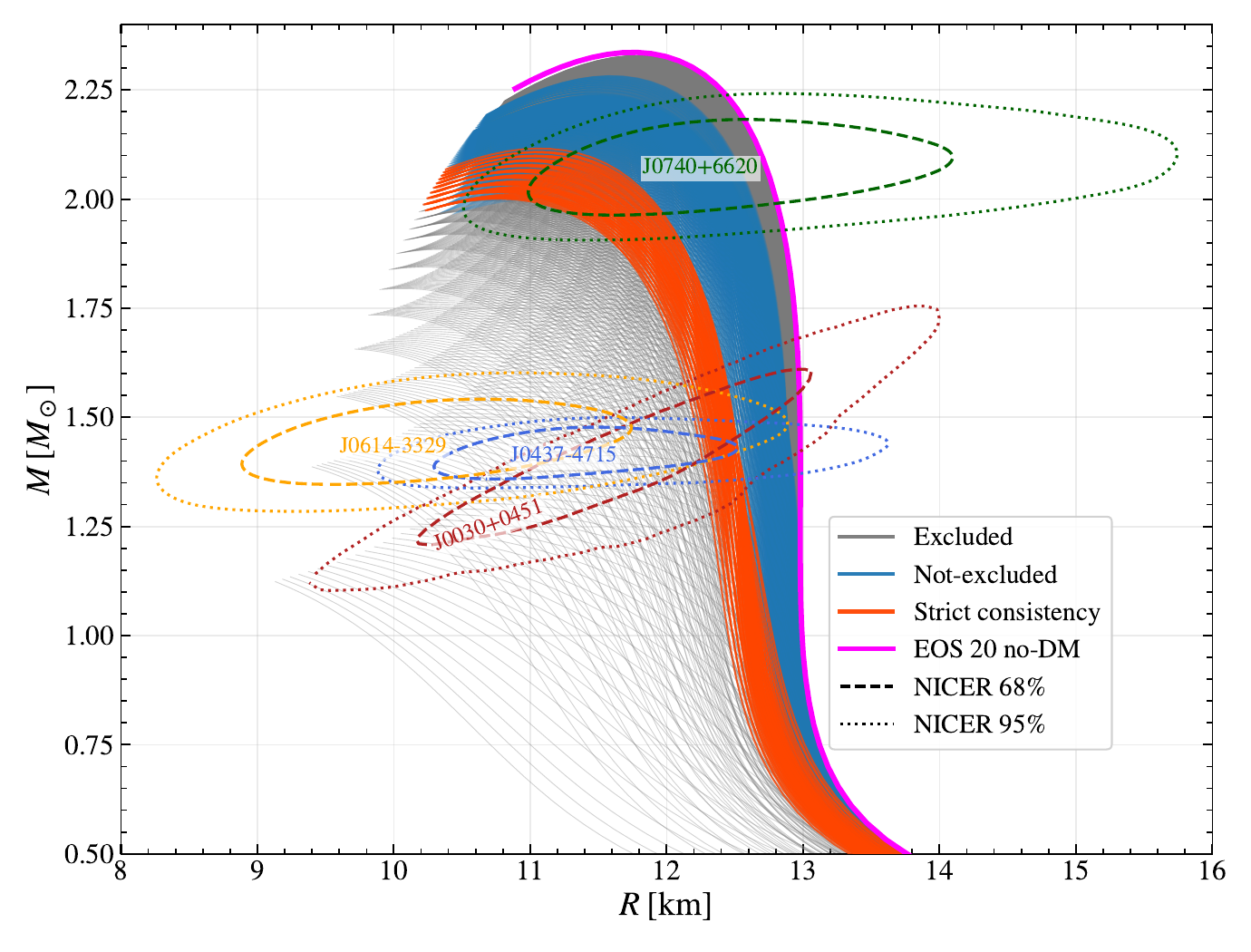}
\par\smallskip
{\small (b) X-PSI/Amsterdam}
\end{minipage}
\caption{
Illustration of the mass--radius classification for the baseline EOS~20 using the Illinois--Maryland and X-PSI/Amsterdam NICER inferences. In both panels, gray curves are excluded by the corresponding conservative basic+all-95\% analysis, blue curves are not excluded by that analysis, and the magenta curve is the no-dark matter EOS~20 sequence. In panel (a), all blue sequences also satisfy the stricter Illinois--Maryland 68\% consistency requirements, and therefore no additional subset appears. In panel (b), orange curves identify the smaller subset that also satisfies the stricter X-PSI/Amsterdam consistency test. Dashed and dotted contours denote the corresponding NICER 68\% and 95\% regions. 
}
\label{fig:eos20_mr_constraint_comparison}
\end{figure*}

To illustrate how the parameter-space classifications discussed above arise from the mass--radius constraints, Fig.~\ref{fig:eos20_mr_constraint_comparison} shows the corresponding stellar sequences for EOS~20 as an illustrative example. EOS~20 is chosen because its purely nucleonic
mass--radius sequence lies outside the X-PSI/Amsterdam
95\% credible region of the compact source
PSR J0614--3329, while neutron-to-dark-fermion
conversion can shift the sequence toward smaller radii
and improve its consistency with this constraint.
Indeed, the dark matter-induced changes in the stellar sequences
and the differences between the observational
classifications are particularly clear for this EOS. The figure does not indicate an additional constraint or independent result, but instead provides a mass--radius representation of the classifications already obtained in the $(m_\chi,G_\chi)$ plane. In both panels, the gray curves show the sequences excluded by the corresponding conservative basic+all-95\% analysis, the blue curves show the sequences that are not excluded by that analysis, and the magenta curve shows the no-dark matter EOS~20 sequence. Orange curves identify the subset of the blue sequences that also satisfies the corresponding stricter consistency test. In the Illinois--Maryland panel, every blue sequence also satisfies the stricter 68\% consistency requirements; therefore, no separate orange subset appears. In the X-PSI/Amsterdam panel, only part of the blue ensemble satisfies the stricter mass--radius requirements and is highlighted in orange.

\section{Soft limiting equations of state}
\label{sec:soft_results}

The two soft limiting EOSs, EOS~6 and EOS~7, provide a stringent test of how much additional softening induced by neutron-to-dark-fermion conversion can be tolerated without violating the maximum-mass constraint. We therefore use these EOSs to determine the conservative excluded and not-excluded regions at this limiting edge for each NICER inference set.

\subsection{Illinois--Maryland analysis}

\begin{figure}[t]
\centering
\includegraphics[width=\columnwidth]{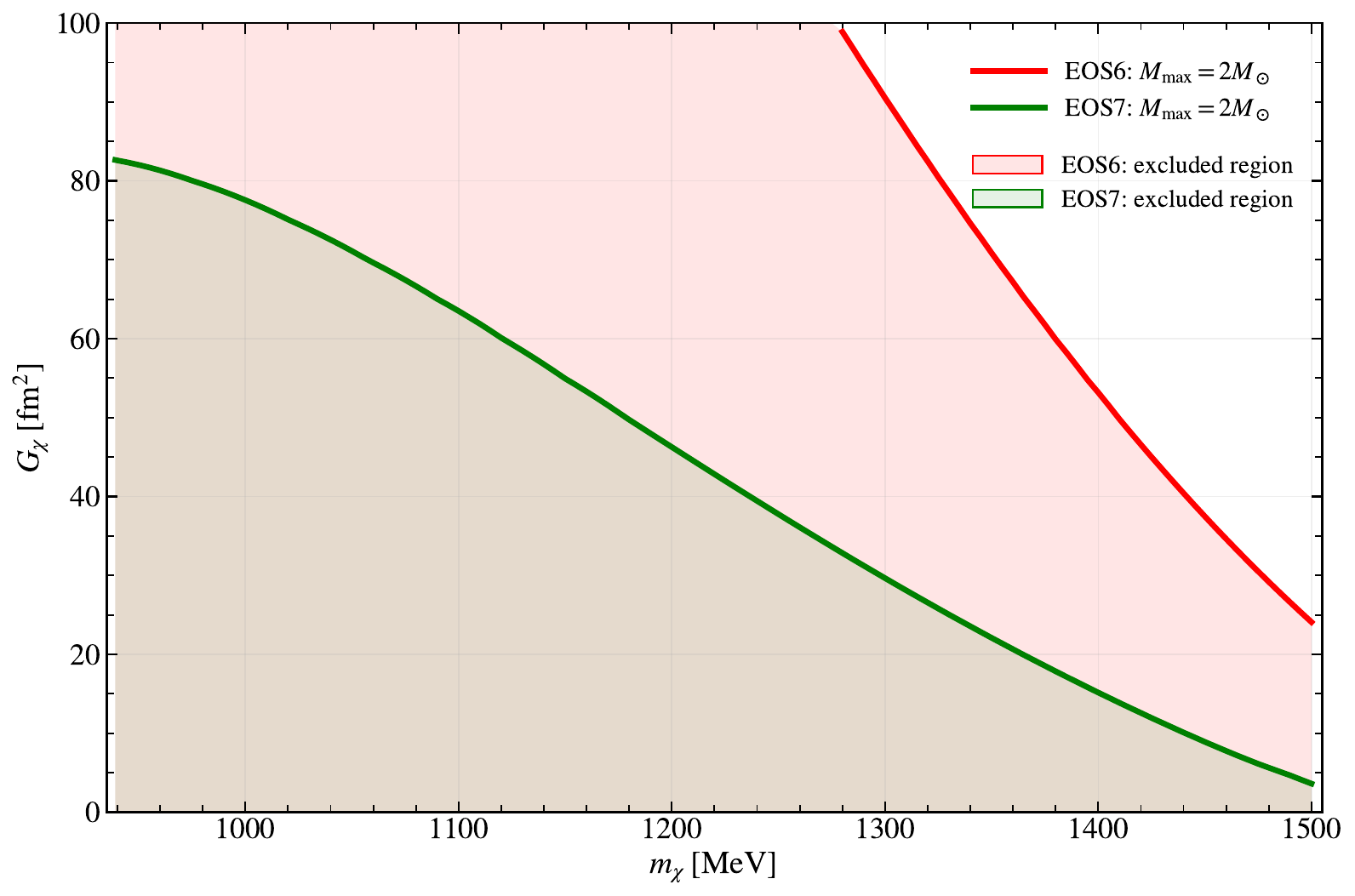}
\caption{
Conservative exclusion boundaries for the soft limiting EOSs, EOS~6 and EOS~7. The same boundaries are obtained under the basic+all-95\% analyses using either the Illinois--Maryland or X-PSI/Amsterdam NICER inferences. For both EOSs, the only active boundary is $M_{\max}=2M_\odot$; the tidal-deformability bound and the corresponding 95\% NICER regions do not add further exclusions over the scanned region. 
}
\label{fig:soft_eos_boundaries}
\end{figure}

Figure~\ref{fig:soft_eos_boundaries} shows the conservative soft-EOS exclusion structure after applying the basic constraints together with all Illinois--Maryland 95\% NICER mass--radius regions. For both EOS~6 and EOS~7, the active conservative boundary is the maximum-mass requirement, $M_{\max}\geq2M_\odot$. The tidal-deformability bound and the Illinois--Maryland 95\% regions are included in the classification, but they do not add further exclusions over the scanned region. This behavior is expected: these soft EOSs are already compact enough to satisfy the adopted tidal-deformability and mass--radius constraints, while the main risk after adding the dark component is that the EOS becomes too soft to support a $2M_\odot$ star.

For the soft subset, the common excluded region, $N_{\rm excl}=2$, is identical to the EOS~7 excluded region because it is contained inside the larger EOS~6 excluded region. It has the minimum--maximum extents
$$
m_\chi=939\text{--}1500~{\rm MeV},
\qquad
G_\chi=1\text{--}80~{\rm fm}^2 .
$$
Equivalently, this region lies below the EOS~7 $M_{\max}=2M_\odot$ boundary.

The common not-excluded region, $N_{\rm excl}=0$, lies above the more restrictive EOS~6 boundary and has the minimum--maximum extents
$$
m_\chi=1280\text{--}1500~{\rm MeV},
\qquad
G_\chi=25\text{--}100~{\rm fm}^2 .
$$
 Thus, if the nucleonic baseline is already close to the maximum-mass limit, avoiding exclusion requires either delaying the dark onset with larger $m_\chi$ or supporting the dark component with stronger repulsion. The quoted intervals are the minimum--maximum extents of the corresponding grid points and should not be interpreted
as indicating that every point within the rectangular
region defined by these limits is or is not excluded.

\subsection{X-PSI/Amsterdam analysis}

Repeating the conservative analysis with the X-PSI/Amsterdam NICER inferences yields the same exclusion structure for the two soft EOSs. The four X-PSI/Amsterdam 95\% mass--radius regions introduce no additional exclusions beyond the maximum-mass condition, while the tidal-deformability bound also remains inactive. Consequently, Fig.~\ref{fig:soft_eos_boundaries} also represents the final conservative X-PSI/Amsterdam result, and the common excluded and not-excluded regions coincide with those obtained using the Illinois--Maryland inferences.

Among the NICER sources, the high-mass pulsar PSR~J0740+6620 is potentially the most relevant for these soft EOSs because it probes the part of the stellar sequence where their limited maximum-mass support becomes important. At the conservative 95\% level, however, the mass--radius regions from both inference sets are sufficiently broad that they do not impose an additional exclusion. At the 68\% level, the PSR~J0740+6620 regions from both inference sets would provide a tighter consistency selection for the soft EOSs, with the more compact X-PSI/Amsterdam region being the more restrictive of the two. These tighter requirements are not used to define exclusions in the present analysis of the soft limiting EOSs.

\section{Stiff limiting equations of state}
\label{sec:stiff_results}

The stiff limiting set, EOS~1--EOS~5 and EOS~21, probes the opposite end of the nucleonic uncertainty range. In the no-dark matter limit, all six EOSs violate the adopted upper bound on $\Lambda_{1.4}$. For this group, neutron-to-dark-fermion conversion can improve compatibility with the tidal-deformability bound and with mass--radius constraints favoring more compact stars.

\subsection{Illinois--Maryland analysis}

\begin{figure}[t]
    \centering
    \includegraphics[width=\columnwidth]{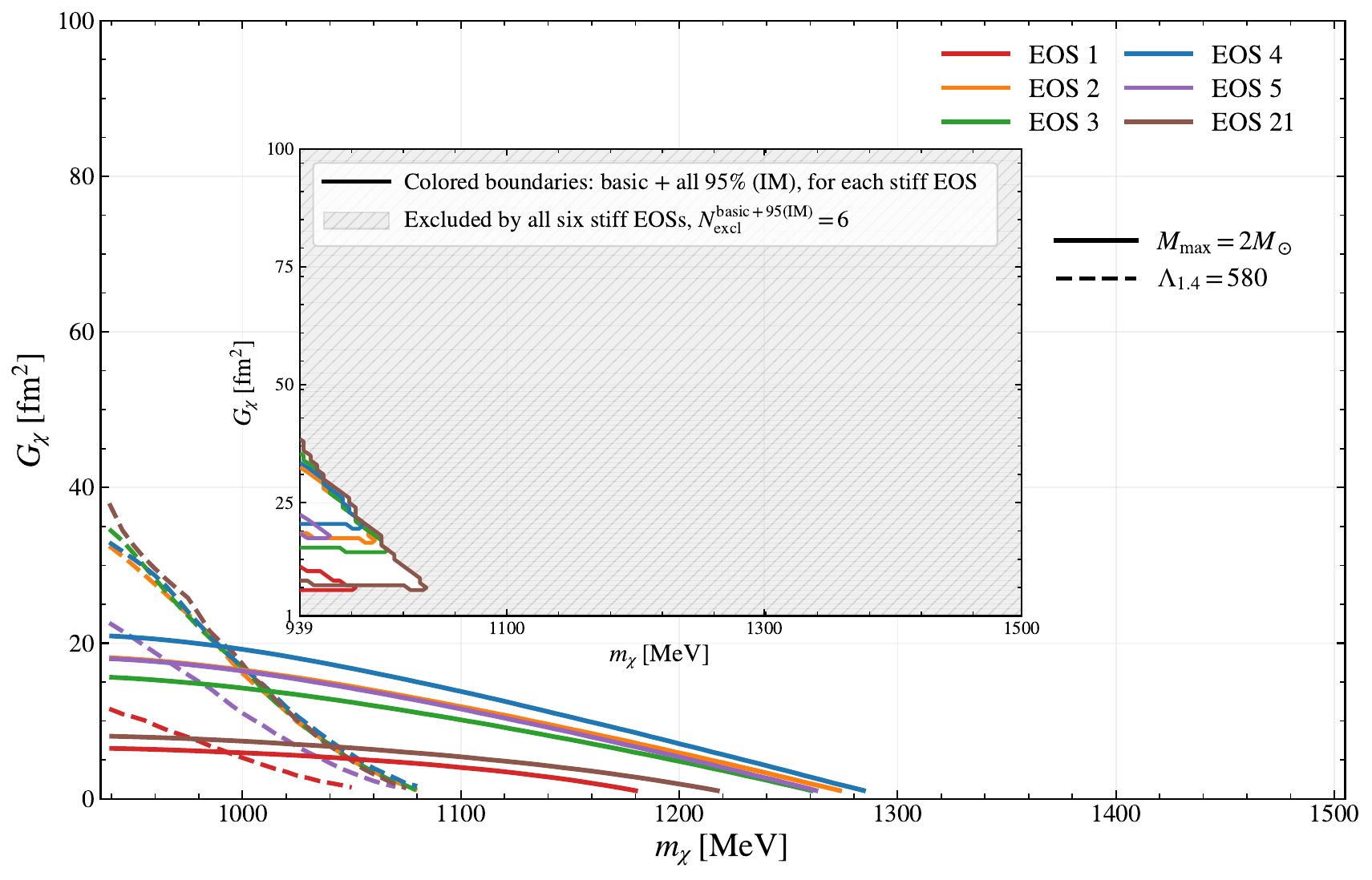}
    \caption{
Conservative exclusion structure for the six stiff limiting EOSs under the Illinois--Maryland analysis. Solid curves show the $M_{\max}=2M_\odot$ boundaries, while dashed curves show the $\Lambda_{1.4}=580$ boundaries; for each EOS, the not-excluded window lies between the corresponding boundaries. Because the Illinois--Maryland 95\% mass--radius regions introduce no additional exclusion, these boundaries coincide with those obtained from the basic constraints. The inset shows the individual not-excluded regions together with the common excluded region, $N_{\rm excl}=6$, indicated by gray shading.  
    }
    \label{fig:stiff_basic_constraints}
\end{figure}

For the stiff limiting EOSs, the Illinois--Maryland 95\% mass--radius regions introduce no additional exclusion beyond the basic constraints. Consequently, the conservative Illinois--Maryland result is determined entirely by the competition between the maximum-mass and tidal-deformability requirements, as shown in Fig.~\ref{fig:stiff_basic_constraints}.

The solid curves in Fig.~\ref{fig:stiff_basic_constraints} show the $M_{\max}=2M_\odot$ boundaries. Below these curves, the conversion is too strong, or the dark self-repulsion is too weak, and the stellar sequence fails the massive-star requirement. The dashed curves show the $\Lambda_{1.4}=580$ boundaries. Above these curves, the dark-sector modification is too weak to reduce the tidal deformability below the adopted upper bound. Thus, for each stiff EOS, the not-excluded window lies between the corresponding solid and dashed boundaries, where the dark component is strong enough to reduce $\Lambda_{1.4}$ without lowering $M_{\max}$ below $2M_\odot$.

Table~\ref{tab:stiff_basic_surviving} summarizes the resulting not-excluded windows for the individual stiff EOSs. These windows are concentrated near the low-$m_\chi$ side of the scan and at moderate values of $G_\chi$. EOS~21 has the largest region, while EOS~5 has the smallest. However, as shown in the inset of Fig.~\ref{fig:stiff_basic_constraints}, the individual windows have no common overlap; that is, no parameter point is simultaneously not excluded for all six stiff EOSs. The gray-shaded region in the inset instead identifies the common excluded domain, $N_{\rm excl}=6$, where the same parameter points are excluded by all six stiff EOSs.

The geometry of this common excluded region is also informative. It covers the full high-$m_\chi$ part of the scan,
$$
m_\chi\geq 1040~{\rm MeV},
$$
for all values of $G_\chi$. It also covers the low- and high-repulsion edges,
$$
G_\chi\leq 6~{\rm fm}^2
\qquad\text{and}\qquad
G_\chi\geq 39~{\rm fm}^2,
$$
for all values of $m_\chi$. Therefore, the only region in which at least one stiff EOS can satisfy the conservative Illinois--Maryland criterion is restricted to
$$
939\leq m_\chi\leq 1035~{\rm MeV},
\qquad
7\leq G_\chi\leq 38~{\rm fm}^2.
$$
These intervals give only the minimum--maximum extent of the relevant grid points and should not be interpreted as a rectangular region or as a common not-excluded region for all six stiff EOSs. In fact, it only means that at least
one of the six stiff EOSs can satisfy the conservative Illinois--Maryland criterion
somewhere inside this range.

\begin{table}[t]
\centering
\caption{
Not-excluded parameter ranges for the stiff limiting EOSs under the conservative basic+all-95\% Illinois--Maryland analysis. Because the Illinois--Maryland 95\% mass--radius regions introduce no additional exclusion, these ranges coincide with the basic-compatible windows determined by $M_{\max}\geq2M_\odot$ and $70\leq\Lambda_{1.4}\leq580$.
}
\label{tab:stiff_basic_surviving}
\begin{ruledtabular}
\begin{tabular}{lcc}
EOS & $m_\chi~[\mathrm{MeV}]$ & $G_\chi~[\mathrm{fm}^2]$ \\
\hline
EOS~1 & 939--980  & 7--11  \\
EOS~2 & 939--995  & 17--32 \\
EOS~3 & 939--1005 & 15--35 \\
EOS~4 & 939--985  & 20--33 \\
EOS~5 & 939--960  & 18--22 \\
EOS~21 & 939--1035 & 7--38  \\
\end{tabular}
\end{ruledtabular}
\end{table}

\subsection{X-PSI/Amsterdam analysis}

\begin{figure}[t]
    \centering
    \includegraphics[width=\columnwidth]{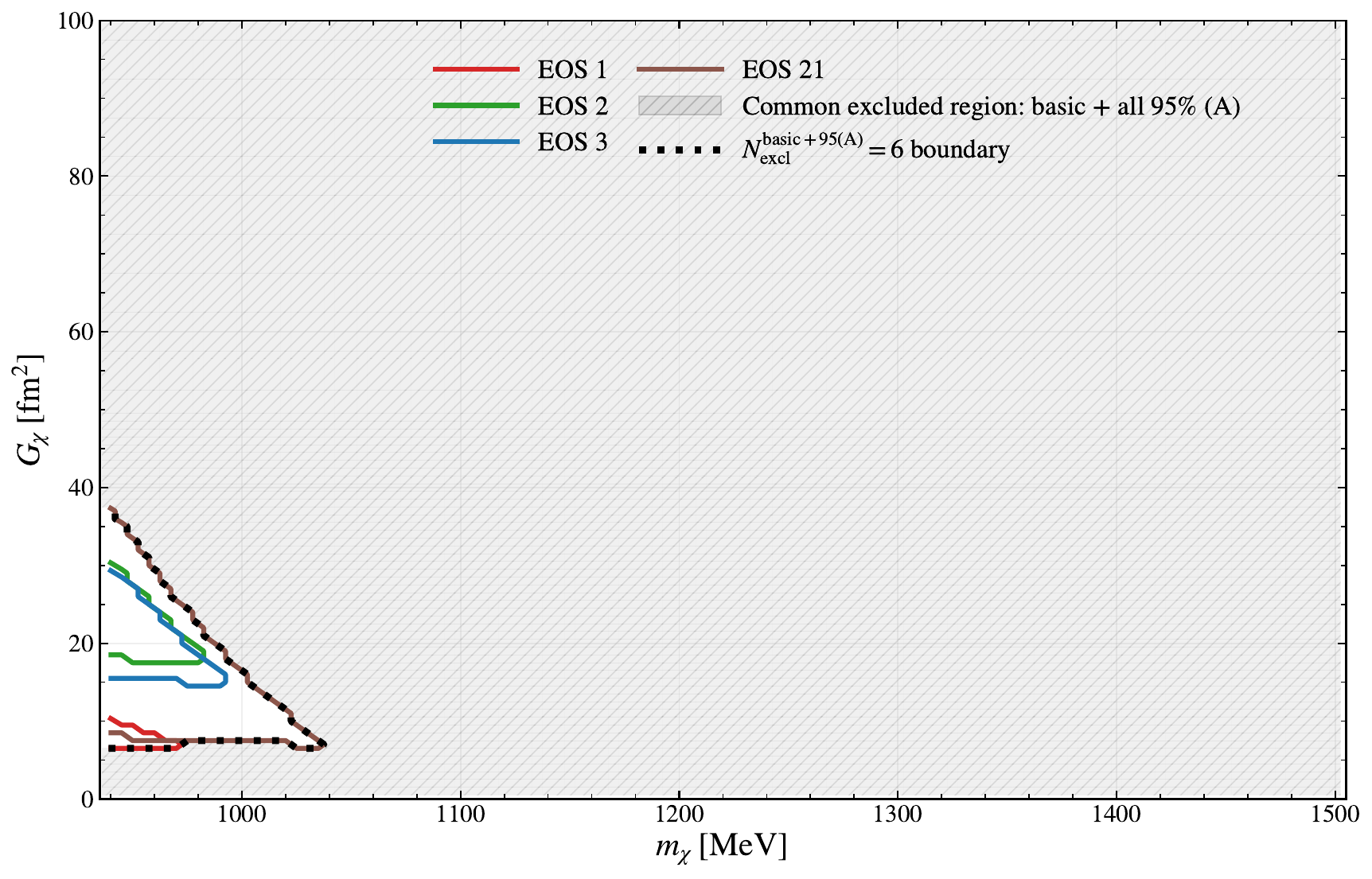}
    \caption{
Stiff-EOS not-excluded boundaries after adding all X-PSI/Amsterdam 95\% mass--radius regions to the basic constraints. EOS~4 and EOS~5 are excluded throughout the scanned parameter space. The colored curves show the remaining not-excluded regions for EOS~1, EOS~2, EOS~3, and EOS~21. The dotted black curve marks the $N_{\rm excl}^{\rm basic+95(A)}=6$ boundary, and the gray-hatched region outside this boundary is the common excluded region, where every parameter point is excluded by all six stiff EOSs.
    }
    \label{fig:stiff_final_95_strict}
\end{figure}

Adding all X-PSI/Amsterdam 95\% mass--radius regions to the basic constraints further reduces the individual stiff-EOS regions. The final not-excluded regions under the conservative basic+all-95\% X-PSI/Amsterdam analysis are shown in Fig.~\ref{fig:stiff_final_95_strict}.  EOS~4 and EOS~5 are fully excluded throughout the scanned parameter space.   Among the remaining stiff EOSs, EOS~1, EOS~2, and EOS~3 retain only narrow not-excluded regions, while EOS~21 has the largest remaining region. Their corresponding parameter ranges are
\[
\begin{aligned}
{\rm EOS~1}:&\quad m_\chi=939\text{--}970~{\rm MeV},\quad G_\chi=7\text{--}10~{\rm fm}^2,\\
{\rm EOS~2}:&\quad m_\chi=939\text{--}980~{\rm MeV},\quad G_\chi=18\text{--}30~{\rm fm}^2,\\
{\rm EOS~3}:&\quad m_\chi=939\text{--}990~{\rm MeV},\quad G_\chi=15\text{--}29~{\rm fm}^2,\\
{\rm EOS~21}:&\quad m_\chi=939\text{--}1035~{\rm MeV},\quad G_\chi=7\text{--}37~{\rm fm}^2.
\end{aligned}
\]
The additional 95\% NICER reduction is driven by the compact PSR J0614--3329 region; the other X-PSI/Amsterdam 95\% regions do not further reduce the basic-compatible stiff-EOS parameter space. The colored curves in Fig.~\ref{fig:stiff_final_95_strict} show the final EOS-by-EOS not-excluded regions. The dotted black curve marks the boundary of the common conservative exclusion, $N_{\rm excl}^{\rm basic+95(A)}=6$. Every parameter point in the gray-hatched region outside this boundary is excluded by all six stiff EOSs.  This common excluded region differs only slightly from the basic and Illinois--Maryland results shown
in Fig.~\ref{fig:stiff_basic_constraints}, with the high-$G_\chi$ boundary shifting from $39$ to $38~\mathrm{fm}^{2}$, while $m_\chi\geq1040~\mathrm{MeV}$ remains excluded for all $G_\chi$. 

At the 68\% consistency level, the Illinois--Maryland regions introduce no further reduction, whereas the X-PSI/Amsterdam regions are considerably more restrictive: among the six stiff EOSs, only EOS~21 retains a small consistent region. These tighter requirements are not used to define exclusions for the stiff limiting set.

\section{EOS-overlapped excluded regions for the full 21-EOS ensemble}
\label{sec:full_ensemble_results}

Finally, we combine the main-baseline, soft, and stiff subsets and count exclusions over the complete 21-EOS ensemble. The fully EOS-overlapped excluded region, $N_{\rm excl}^{21}=21$, provides the strongest conservative exclusion within the adopted ensemble, because every EOS in the full set excludes the same region of the dark-sector parameter space. This should not be confused with an analysis of a common not-excluded region. Points outside the common excluded region do not necessarily satisfy the constraints for all EOSs; instead, their status can still depend on the chosen nucleonic EOS.

\begin{figure}[t]
    \centering
    \includegraphics[width=\columnwidth]{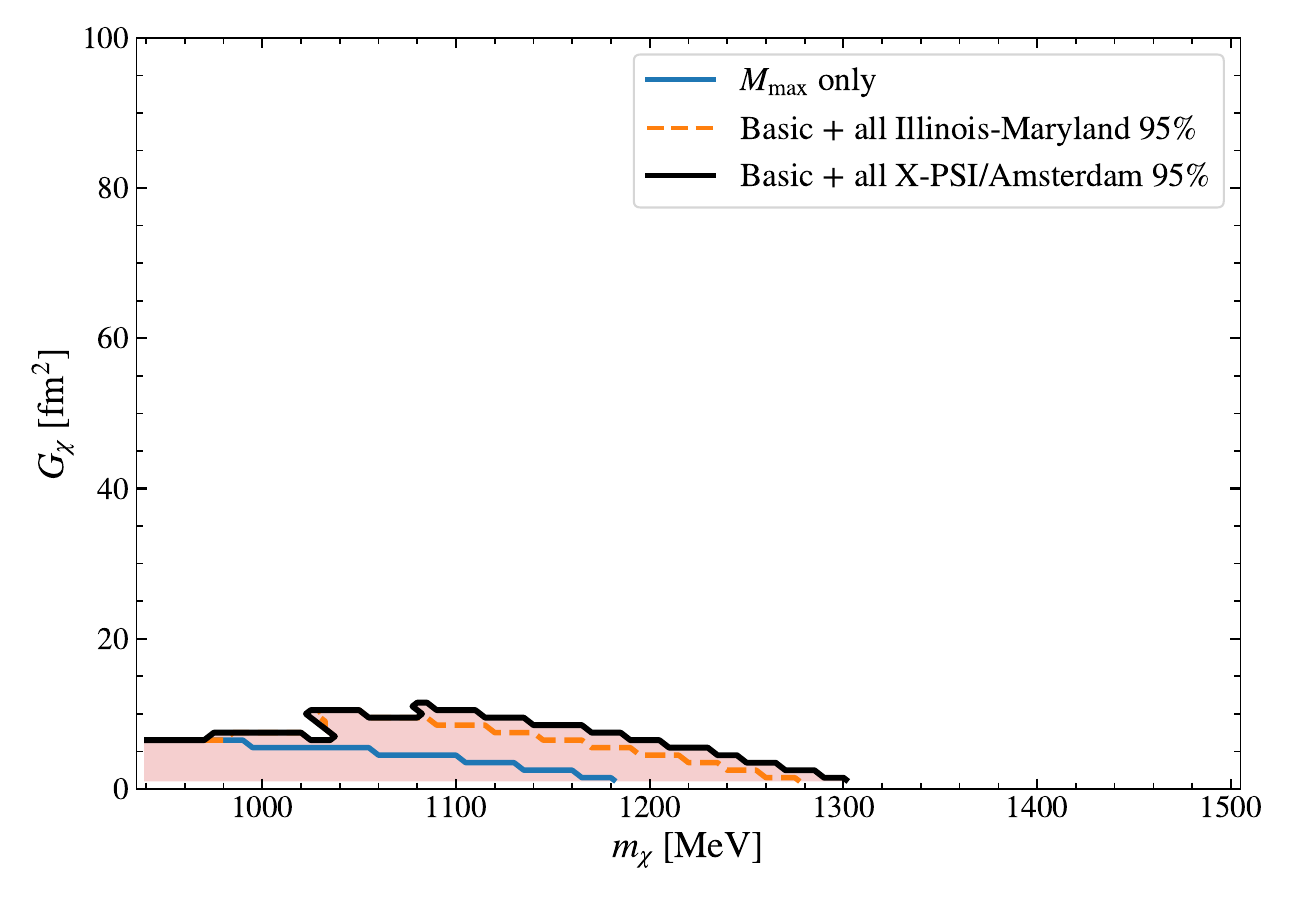}
    \caption{
Fully EOS-overlapped excluded regions for the complete 21-EOS ensemble. The blue curve shows the common exclusion obtained from the $M_{\max}\geq2M_\odot$ condition alone. The orange dashed curve shows the conservative basic+all-95\% Illinois--Maryland result and coincides with the basic-constraint boundary because the Illinois--Maryland 95\% mass--radius regions introduce no additional exclusion. The black curve shows the corresponding conservative basic+all-95\% X-PSI/Amsterdam result, and the shaded region marks its common excluded domain.
    }
    \label{fig:all21_common_excluded_boundaries}
\end{figure}

Figure~\ref{fig:all21_common_excluded_boundaries} compares the common excluded boundaries obtained from the maximum-mass condition alone and from the two conservative NICER analyses. The maximum-mass condition alone gives a common excluded region for all 21 EOSs with minimum--maximum extent
\[
m_\chi=939\text{--}1180~{\rm MeV},
\qquad
G_\chi=1\text{--}6~{\rm fm}^2 .
\]

Adding the tidal-deformability bound gives the full basic constraint set. The Illinois--Maryland 95\% mass--radius regions introduce no additional exclusion beyond this basic set. Therefore, the orange dashed boundary in Fig.~\ref{fig:all21_common_excluded_boundaries} represents both the basic-constraint boundary and the conservative basic+all-95\% Illinois--Maryland result, which has minimum--maximum extent
\[
m_\chi=939\text{--}1275~{\rm MeV},
\qquad
G_\chi=1\text{--}10~{\rm fm}^2 .
\]
The enlargement of this region relative to the $M_{\max}$-only result is driven by the tidal-deformability requirement, although the Illinois--Maryland 95\% mass--radius regions are also included in the analysis.

Starting from the same basic constraint set, the independent conservative X-PSI/Amsterdam analysis gives a slightly larger common excluded region. After applying all X-PSI/Amsterdam 95\% mass--radius regions, its minimum--maximum extent becomes
\[
m_\chi=939\text{--}1300~{\rm MeV},
\qquad
G_\chi=1\text{--}11~{\rm fm}^2 .
\]
The two NICER inference sets are therefore applied independently to the same basic constraints.

The resulting EOS-overlapped exclusions are characterized primarily by low values of $G_\chi$ within the corresponding $m_\chi$ ranges, where the repulsive dark self-interaction is insufficient to counteract the EOS softening induced by neutron-to-dark-fermion conversion. As before, the quoted intervals denote the minimum--maximum extents of the common excluded grid points and should not be interpreted as rectangular exclusion regions.

For completeness, Table~\ref{tab:common_regions_summary} provides a
compact summary of the common excluded and common not-excluded regions
for all four EOS groups. The two NICER analyses are listed separately
when they produce distinct results, while identical or nearly identical
results are reported together to avoid repetition.

\begin{table*}[t]
\caption{
Summary of the common excluded and common not-excluded regions for the
different nucleonic EOS subsets under the conservative basic+all-95\%
Illinois--Maryland and X-PSI/Amsterdam analyses. The two NICER inference
sets are applied independently to the same basic constraints. The
Illinois--Maryland 95\% regions introduce no additional exclusion, so
the corresponding results coincide with those obtained from the basic
constraints.  For entries
expressed as intervals, the quoted ranges denote the minimum--maximum
extents of the corresponding grid points and should not be interpreted
as rectangular regions in the $(m_\chi,G_\chi)$ plane. However, entries referring to the full $m_\chi$ or $G_\chi$ scan mean that every scanned value of that parameter belongs to the corresponding region, rather than giving a minimum--maximum extent.
}
\label{tab:common_regions_summary}

\begin{ruledtabular}
\begin{tabular}{
p{0.14\textwidth}
p{0.17\textwidth}
p{0.35\textwidth}
p{0.20\textwidth}
}

\textbf{EOS subset}
&
\textbf{NICER inference}
&
\makebox[\linewidth][c]{\shortstack{\footnotesize\textbf{Common excluded}\\[-0.3ex]
\footnotesize\textbf{region}}}
&
\makebox[\linewidth][c]{\shortstack{\footnotesize\textbf{Common not-excluded}\\[-0.3ex]
\footnotesize\textbf{region}}}
\\
\hline

13 main-baseline EOSs
&
Illinois--Maryland
&
\mbox{$m_\chi=939\text{--}1275~\mathrm{MeV}$},
\quad
\mbox{$G_\chi=1\text{--}12~\mathrm{fm}^{2}$}
&
For all scanned $m_\chi$:
\mbox{$G_\chi\geq36~\mathrm{fm}^{2}$};
\newline
for all scanned $G_\chi$:
\mbox{$m_\chi\geq1395~\mathrm{MeV}$}
\\[1.2ex]

&
X-PSI/Amsterdam
&
\mbox{$m_\chi=939\text{--}1300~\mathrm{MeV}$},
\quad
\mbox{$G_\chi=1\text{--}12~\mathrm{fm}^{2}$}
&
\mbox{$m_\chi=939\text{--}1010~\mathrm{MeV}$}
\newline
\mbox{$G_\chi=33\text{--}80~\mathrm{fm}^{2}$}
\\

\noalign{\vskip 1.8ex}

Soft EOS subset
&
Both inference sets
&
\mbox{$m_\chi=939\text{--}1500~\mathrm{MeV}$},
\quad
\mbox{$G_\chi=1\text{--}80~\mathrm{fm}^{2}$}
&
\mbox{$m_\chi=1280\text{--}1500~\mathrm{MeV}$}
\newline
\mbox{$G_\chi=25\text{--}100~\mathrm{fm}^{2}$}
\\

\noalign{\vskip 1.8ex}

Stiff EOS subset
&
Both inference sets
&
For all scanned $m_\chi$:
\mbox{$G_\chi\leq6~\mathrm{fm}^{2}$} or
\mbox{$G_\chi\gtrsim39~\mathrm{fm}^{2}$};
for all scanned $G_\chi$:
\mbox{$m_\chi\geq1040~\mathrm{MeV}$}
&
No common region
\\

\noalign{\vskip 1.8ex}

Full 21-EOS ensemble
&
Illinois--Maryland
&
\mbox{$m_\chi=939\text{--}1275~\mathrm{MeV}$},
\quad
\mbox{$G_\chi=1\text{--}10~\mathrm{fm}^{2}$}
&
No common region
\\[1.2ex]

&
X-PSI/Amsterdam
&
\mbox{$m_\chi=939\text{--}1300~\mathrm{MeV}$},
\quad
\mbox{$G_\chi=1\text{--}11~\mathrm{fm}^{2}$}
&
No common region
\\

\end{tabular}
\end{ruledtabular}
\end{table*}

\section{Summary and conclusions}
\label{sec:conclusions}

We have studied neutron-to-dark-fermion conversion in cold neutron star matter in the single-$\chi$ channel, imposing $\mu_\chi=\mu_n$ and including a repulsive dark self-interaction through the effective coupling $G_\chi$. The scan covers $m_\chi=939$--$1500~\mathrm{MeV}$ and $G_\chi=1$--$100~{\rm fm}^2$, so it includes both the near-threshold regime close to the neutron mass and heavier in-medium dark states that can appear only when the neutron chemical potential in the stellar core becomes sufficiently large.

The main physical trend is governed by the competition between the dark-particle mass and its repulsive self-interaction. Smaller $m_\chi$ and $G_\chi$ favor a larger dark-fermion population leading to stronger softening, depending on the nucleonic EOS, which reduces the maximum mass and can make more compact stellar configurations.  Increasing $m_\chi$ delays the onset of the dark component, while increasing $G_\chi$ raises the energetic cost of populating it. Both effects drive the EOS and the mass--radius and tidal-deformability sequences back toward the purely nucleonic (no-dark matter) limit. The quantitative response, however, depends strongly on the nucleonic EOS, which motivates the EOS-by-EOS and EOS-overlap analyses carried out in this work.

For the 13 main-baseline EOSs, we applied the two conservative analyses separately and on equal footing by combining the basic constraints with all four Illinois--Maryland $95\%$ regions and, independently, with all four X-PSI/Amsterdam $95\%$ regions. The Illinois--Maryland regions introduce no additional exclusion beyond the basic constraints. Because the tidal-deformability bound  introduces
no additional exclusions for this EOS group, the corresponding boundaries coincide with those determined by the maximum-mass condition. The Illinois--Maryland common excluded region spans $m_\chi=939$--$1275~\mathrm{MeV}$ and $G_\chi=1$--$12~\mathrm{fm}^2$ while its common not-excluded region covers the full scanned $m_\chi$ range for $G_\chi\geq36~\mathrm{fm}^2$ and the full scanned $G_\chi$ range for $m_\chi\geq1395~\mathrm{MeV}$. For the X-PSI/Amsterdam analysis, the common excluded region spans $m_\chi=939$--$1300~\mathrm{MeV}$ and $G_\chi=1$--$12~\mathrm{fm}^2$, whereas the common not-excluded region spans $m_\chi=939$--$1010~\mathrm{MeV}$ and $G_\chi=33$--$80~\mathrm{fm}^2$.

The detailed stricter consistency tests are applied only to the 13 main-baseline EOSs and are not used as conservative exclusion criteria. The Illinois--Maryland $68\%$ regions produce no further reduction of the corresponding conservatively not-excluded parameter space. The stricter X-PSI/Amsterdam test, using the $68\%$ regions of PSR~J0030+0451, PSR~J0740+6620, and PSR~J0437--4715 together with the $95\%$ region of PSR~J0614--3329, does not yield a parameter point common to all 13 EOSs, but identifies a small region consistent with 11 of the 13 baseline EOSs at $m_\chi=939$--$950~\mathrm{MeV}$ and $G_\chi=19$--$21~\mathrm{fm}^2$. These intervals denote only the minimum--maximum extents of the corresponding grid points and should not be interpreted as rectangular regions in the $(m_\chi,G_\chi)$ plane.

The soft and stiff limiting EOSs further show the residual sensitivity of the constraints to the nucleonic stiffness. For the soft EOSs, the active conservative constraint is the maximum-mass requirement, and the Illinois--Maryland and X-PSI/Amsterdam $95\%$ regions give the same conservative result. Avoiding exclusion for both EOS~6 and EOS~7 shifts the remaining parameter space toward heavier $m_\chi$ and stronger dark self-repulsion. For the stiff EOSs, the competition is different: the dark component can help reduce the large tidal deformability of the no-dark matter sequences without lowering $M_{\max}$ below $2M_\odot$. The two conservative NICER analyses give nearly identical common exclusions, differing only by one grid step at high $G_\chi$. Individual stiff EOSs can satisfy this balance only in narrow windows at low $m_\chi$, close to the lower edge of the scan near the neutron mass, and intermediate $G_\chi$. However, no parameter point satisfying the conservative constraints is common to all six stiff limiting EOSs.

Finally, combining all 21 EOSs gives the strongest EOS-overlapped exclusion, identifying the parameter points that are excluded for every EOS in the full set. Under the conservative Illinois--Maryland analysis, the common excluded region spans $m_\chi=939$--$1275~\mathrm{MeV}$ and $G_\chi=1$--$10~\mathrm{fm}^2$ in minimum--maximum extent. This region is identical to that obtained from the basic constraints because the Illinois--Maryland $95\%$ regions introduce no additional exclusion; its enlargement relative to the $M_{\max}$-only result is driven by the tidal-deformability requirement. Under the conservative X-PSI/Amsterdam analysis, the common excluded region spans $m_\chi=939$--$1300~\mathrm{MeV}$ and $G_\chi=1$--$11~\mathrm{fm}^2$. Therefore, within the present dark-sector model, observational set, and calibrated RMF ensemble, both independent NICER analyses exclude the low-$G_\chi$ part of the dark-sector parameter space over a relatively broad range of $m_\chi$,  for every one of the 21 EOS parametrizations considered.  By contrast, the remaining not-excluded parameter space is more EOS dependent.

An important general lesson of our analysis is that
neutron-to-dark-fermion conversion at low $m_\chi$ and
weak $G_\chi$ can lead to substantial EOS softening,
challenging the requirement that neutron stars support
masses of at least $2M_\odot$. The fact that this behavior
is found across the soft, main-baseline, and stiff
nucleonic EOSs considered here suggests that it could
remain an important constraint for the physical
dense-matter EOS, regardless of its microscopic composition.
The precise excluded parameter region, however,
depends on the underlying EOS.

Future improvements can include updated NICER, X-ray, radio, and gravitational-wave constraints, together with complementary observables such as rotating-star properties, radial and non-radial oscillation modes, cooling behavior, and dynamical simulations. These extensions would help distinguish dark-sector effects from ordinary EOS uncertainties. On the theoretical side, an important extension would be to repeat the analysis using EOSs containing additional high-density degrees of freedom, particularly QHC21-type quark--hadron crossover matter \cite{Kojo:2021wax}. A dark component produced in the hadronic regime could modify the thermodynamic input entering the crossover construction, and such an analysis would show how the common exclusions found here change beyond the nucleonic RMF framework considered in this work. It would also be useful to extend the analysis to other conversion channels, including multi-particle dark final states.

\begin{acknowledgments}

D.R.K. was supported by Polish NCN Grant No. 2023/51/B/ST9/02798 and SONATINA 7 Grant No. 2023/48/C/ST2/00297. Part of this work was carried out during D.R.K.'s stay at the Few-body Systems in Physics Laboratory at RIKEN. D.R.K. gratefully acknowledges the kind hospitality of the laboratory, and in particular Prof. Emiko Hiyama and Prof. Makoto Oka, during the completion of this work. D.R.K. thanks Tuhin Malik for fruitful discussions concerning the nucleonic EOSs used in this work. M.S.
acknowledges support from the National Science
Centre, Poland (NCN), through SONATINA 7 Grant
No. 2023/48/C/ST2/00297. M.S. also acknowledges the hospitality of the Few-body Systems in Physics Laboratory at the RIKEN Nishina Center, particularly Prof. Makoto Oka, and thanks Prof. Emiko Hiyama for the kind invitation.

\end{acknowledgments}

\end{document}